\documentclass[reprint,superscriptaddress,nofootinbib,aps,pre]{revtex4-1}
\pdfoutput=1
\usepackage[utf8]{inputenc}
\usepackage[english]{babel}
\usepackage{amsmath}
\usepackage{amsfonts}
\usepackage{amssymb}
\usepackage[hyperindex,colorlinks,hyperfootnotes = false,citecolor = black,]{hyperref}
\usepackage[usenames]{color}
\usepackage{graphicx}

\begin{document}
\title{Emergence of slow-switching assemblies in structured neuronal networks}
\author{Michael T. Schaub}
\email{michael.schaub09@imperial.ac.uk}
\thanks{equal contribution}
\affiliation{Department of Mathematics, Imperial College London, London SW7 2AZ, U.K.}
\author{Yazan N. Billeh}
\email{ybilleh@caltech.edu}
\thanks{equal  contribution}
\affiliation{Computation and Neural Systems Program, California Institute of Technology, Pasadena, California 91125}
\author{Costas A. Anastassiou}
\affiliation{Allen Institute for Brain Science, Seattle, Washington 98103}
\author{Christof Koch}
\affiliation{Allen Institute for Brain Science, Seattle, Washington 98103}
\author{Mauricio Barahona}
\affiliation{Department of Mathematics, Imperial College London, London SW7 2AZ, U.K.}
\date{\today}

\begin{abstract}
Unraveling the interplay between connectivity and spatio-temporal dynamics in
neuronal networks is a key step to advance our understanding of neuronal information processing.
Here we investigate how particular features of network connectivity underpin the propensity of neural networks to generate slow-switching assembly (SSA) dynamics, 
i.e., sustained epochs of increased firing within assemblies of neurons which transition slowly between different assemblies throughout the network. 
We show that the emergence of SSA activity is linked to spectral properties of the asymmetric synaptic weight matrix. 
In particular, the leading eigenvalues that dictate the slow dynamics exhibit a gap with respect to the bulk of the spectrum, and the associated Schur vectors exhibit a measure of block-localization on groups of neurons, thus resulting in coherent dynamical activity on those groups.
Through simple rate models, we gain analytical understanding of the origin and importance of the spectral gap, and use these insights to develop new network topologies with alternative connectivity paradigms which also display SSA activity.
Specifically, SSA dynamics involving excitatory and inhibitory neurons can be achieved by modifying the connectivity patterns between both types of neurons. 
We also show that SSA activity can occur at multiple timescales reflecting a hierarchy in the connectivity, and demonstrate the emergence of SSA in small-world like networks.
Our work provides a step towards understanding how network structure (uncovered through advancements in neuroanatomy and connectomics) can impact on spatio-temporal neural activity and constrain the resulting dynamics. 
\end{abstract}
\maketitle

\section*{Author Summary}
Neural networks display a wide range of spatio-temporal behaviors.
Understanding how this complex orchestration of neuronal firing activity is determined by the structure of the network (\textit{i.e.}, its wiring) is an important step towards comprehending how neural computation is manifested, especially given the growing experimental access to temporal recordings and connectomics.  
Here we investigate the link between network structure and the dynamics of neuronal assemblies in the context of leaky-integrate-and-fire (LIF) networks.
We show how structural features in the wiring of the network can introduce additional time-scales to the dynamics, and how such structured wiring can lead to
spatio-temporally segregated, coherent activity of groups of neurons.
Using rate models we gain insight into how such spatio-temporal dynamics emerge as a direct consequence of the spectral properties of the network, and use this understanding to examine further circuit topologies that enable phenomenologically similar behavior, yet rely on fundamentally different connectivity and functional patterns.

\section{Introduction}

Neuronal ensembles exhibit a broad repertoire of activity patterns. Such dynamics are governed by a time-evolving network of synaptic connections with an intricate, yet structured, organization.
Due to the advancement of connectomics, our knowledge about such networks is rapidly growing, and increasingly detailed maps of neuronal wiring are becoming available.
In parallel, modern recording techniques, such as calcium imaging and multi-electrode arrays, allow neuroscientists to monitor the activity from thousands of neurons simultaneously, with recordings from entire brains at single neuron resolution becoming technologically feasible~\cite{Buzsaki2004,Du2011,Ahrens2013}.
The observed dynamics of neural networks exhibit an interplay of structured spatio-temporal scales, which underpin a wide range of cognitive functions \cite{Buzsaki2010}.

The idea that neuronal group activity induced by network structure is at the core of neural computation dates back at least to the work of Hebb \cite{Hebb1949}, who hypothesized that the transient activity of groups of neurons (so called \textit{cell assemblies}) is the currency of information processing \cite{Harris2005,Buzsaki2010}.
This notion is supported by recent experiments showing that reciprocal connections between neurons occur above chance level \cite{Song2005,Perin2011}, especially if neurons receive common inputs \cite{Yoshimura2005,Otsuka2011}. In the case of the visual system, for instance, excitatory neurons with similar response features tend to be more connected to each other \cite{Ko2011,Harris2013}.
Moreover, studies have demonstrated that neurons exhibit layer-specific connectivities within rodent sensory cortex \cite{Lefort2009} and neocortex \cite{Yassin2010}.
In addition, organized architectures have been observed to occur at multiple hierarchical scales \cite{Shimono2014,McGinley2013,Savic2000,Felleman1991} and in non-mammalian organisms \cite{Ito2013}.
These findings suggest that cortical regions contain well-defined subnetworks. 
However, the underlying question is whether given a network topology, we can predict the potential of the network to sustain structured spatio-temporal activity. 
Such questions are not only of interest for network dynamics, but have also implications for memory formation and learning, since neural networks undergo topological changes over time due to plasticity \cite{Pavlides1988,Hyman2003}.

\begin{figure}[tb!]
  \centering
  \includegraphics{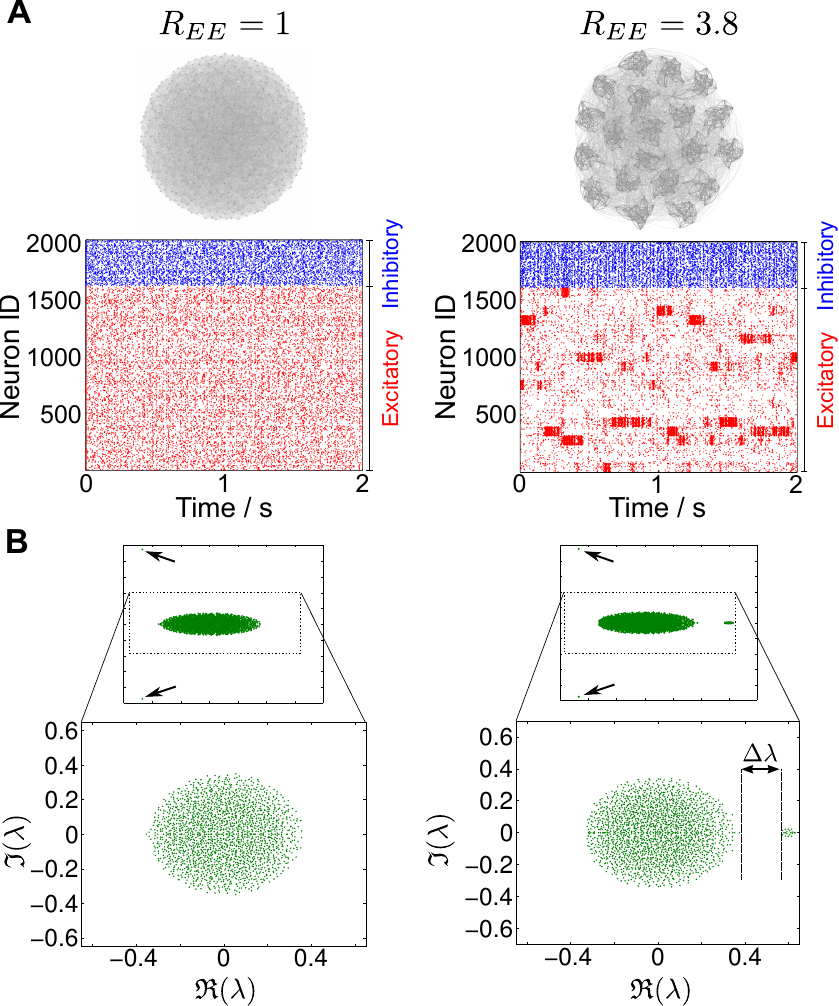}
  \caption{Network dynamics and eigenvalue spectra of two LIF networks: one with uniform 
  synaptic connections (left), and one with 20 groups of clustered excitatory connections (right). 
  To create the clustered network, excitatory neurons were partitioned into groups (with in-group connection probability $p_\text{in}^{EE}$ and out-group connection probability $p_\text{out}^{EE}< p_\text{in}^{EE}$) while keeping the average connectivity constant (see Materials and Methods and Ref.~\cite{Litwin-Kumar2012}).
  The ratio $R_{EE} = p_\text{in}^{EE} / p_\text{out}^{EE}$ controls the strength of the excitatory clustering.
  \textbf{A} Visualization of the network topologies (top) and exemplars of raster plots (bottom). The dynamics of the clustered network exhibit the banded structure associated with slow-switching group activity. 
  The magnitude of the activity can be characterized statistically \textit{a posteriori} from the data through the spike-rate variability metric $\widehat S$, defined in Materials and Methods Eq.~\eqref{eq:S_hat}, as discussed in the text. In this case, the unclustered network has $\widehat S = 0.035$ while the clustered network has 
  a much larger value $\widehat S =8.23$.
   \textbf{B} Eigenvalue spectra of the network weight matrices $W$.
 The weighted connectivity matrix of the clustered network exhibits a clear eigengap $\Delta \lambda$ separating the 19 eigenvalues with largest real parts from the cloud of eigenvalues in the bulk. 
 There is no such eigengap for the unclustered network. As indicated by the two arrows, both matrices have a pair of complex conjugate eigenvalues associated with the (damped) global activation modes of the networks characteristic of balanced networks (see text~and~Ref.~\cite{Murphy2009}).}
  \label{fig:1}
\end{figure}

Recently, it has been shown computationally (see e.g. Ref.~\cite{Litwin-Kumar2012}) that leaky-integrate-and-fire (LIF) networks with equal excitatory and inhibitory connection net strengths (\textit{i.e.} balanced \cite{Vreeswijk1998})  yet with clustered excitatory connections, can exhibit prolonged heightened group activity, with the activity transitioning between groups in the network (Fig.~\ref{fig:1}A).
Here we characterize the emergence of such slow-switching segregated dynamics in balanced LIF networks as a result of the network connectivity.
Specifically, we find that the spectral properties of the synaptic weight matrix  (i.e., the existence of an eigenvalue gap and a block-localized dominant subspace) provide a criterion to predict the appearance of such activity in the network. 
We then use simple linear rate models to gain insight into the mechanisms underpinning the origin of such dynamics in structurally clustered LIF networks. 
Using these insights, we construct novel LIF topologies that display slow-switching group activity
with distinct properties: involving both inhibitory and excitatory neurons; 
exhibiting multiple slow time-scales; 
as well as demonstrating the possibility of such dynamics in networks with no obvious 
clustered connectivity, such as small-worlds.
Finally, we discuss briefly possible implications of the different wiring schemes for neural computation.

\section{Results}

\subsection{Slow-switching assemblies in LIF networks with clustered excitatory connections: spectral insights}

Clustered excitatory topologies in a balanced LIF network can lead to dynamics in which localized high activity states transition between assemblies of neurons within the network \cite{Litwin-Kumar2012}.
This is illustrated in Figure \ref{fig:1}A, where the dynamics of an unstructured and a balanced clustered network with $20$ groups are shown side by side (see Materials and Methods for a description of the networks). 
Hereafter, we will refer to such activity as \textit{slow-switching assembly} (SSA) dynamics. 
Visually, SSA dynamics manifests itself as bands of increased activity in the raster plots, and can be statistically quantified from the resulting spike-train dynamics \textit{a posteriori} (see~Eq.~\eqref{eq:S_hat} in Materials and Methods).
Ideally, however, we would like to establish \textit{a priori}, solely from the given connectivity, the possibility of such dynamical patterns emerging.

The full dynamics of LIF networks are notoriously difficult to analyze due to their inherent non-linearity; hence an exact analytical treatment of the dynamical evolution for an arbitrary clustered topology is essentially intractable. However, two concepts from spectral graph theory and linear systems provide valuable insights: 
(i)~for symmetric, non-negative connectivity matrices, it can be shown that a modular network structure implies a gap in the spectrum of the graph (i.e., in the set of eigenvalues of the weight matrix)~\cite{Luxburg2007,Zhang2013}, as well as the block-localization of the associated eigenvectors on the modules of the network (noting that isolated eigenvalues may also be the result of other features~\cite{Nadakuditi2013});
(ii) for linear systems, a gap in the spectrum of the graph results in a \textit{separation of time scales} in the dynamical process \cite{Simon1961,Galan2008, Murphy2009}.
This relation between the modular structure, the eigenvalues and associated eigenvectors, and linear network dynamics can be used to discover modular structures in graphs from a dynamical perspective \cite{Delvenne2010,Schaub2012,Delvenne2013, Billeh2014}.

In fact, the weighted connectivity matrices of unclustered and clustered LIF networks~\cite{Litwin-Kumar2012} display different spectral characteristics.
Figure \ref{fig:1}B shows the spectra of two networks with 1600 excitatory and 400 inhibitory neurons with unclustered (left) and clustered (right) topologies.
In the unclustered case, we find the expected circular distribution of eigenvalues, which follows from the properties of random graphs~\cite{Rajan2006,Sommers1988}, although the presence of groups of neurons with different cardinalities and variances means that the eigenvalue distribution is not completely uniform on the circle~\cite{Rajan2006}. 
Note also that the balanced construction of the LIF network, with a marginally larger inhibitory input for each neuron in order to keep the network stable, leads to the existence of one pair of complex conjugate eigenvalues which lies separate from the main bulk of the spectrum (black arrows in Figure \ref{fig:1}B).
This eigenvalue pair is associated with the global activation mode of the network (as explained
below in Section~\ref{sec:linear_rates} and in Ref.~\cite{Murphy2009}).
As shown in Figure~\ref{fig:1}A, this unclustered network exhibits the expected asynchronous, unstructured neuronal spiking dynamics.

\begin{figure}[tb!]
  \centering
  \includegraphics{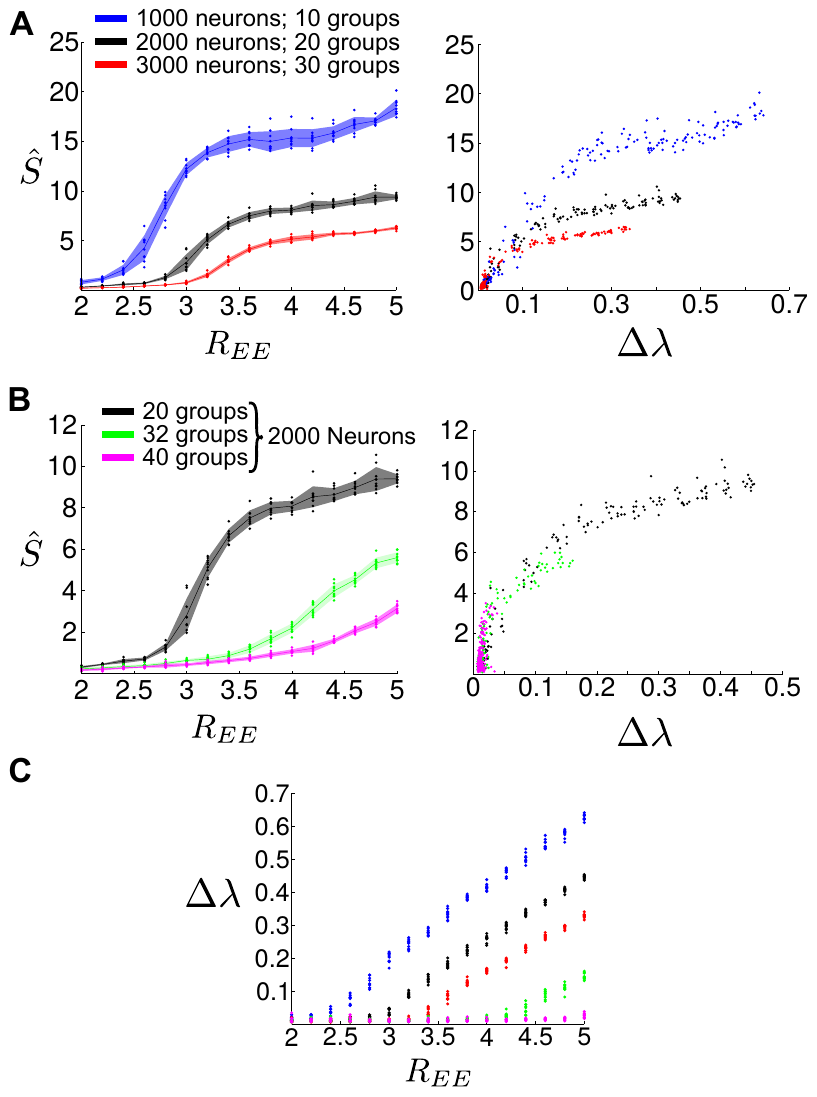}
  \caption{The relationship of observed SSA dynamics with the structural connectivity clustering of the LIF network, $R_{EE}$, and the spectral gap, $\Delta \lambda$.
The presence of SSA dynamics is quantified through the spike-rate variability metric ($\widehat S$), which measures the variance of the firing rates of the assemblies normalized by a randomly shuffled bootstrap: $\widehat S$ increases with increasing SSA activity and $\widehat S \approx 0$ for completely asynchronous activity (as in the unclustered case in Figure~\ref{fig:1}A) (see Materials and Methods). 
 \textbf{A}   Spike-rate variability is plotted as a function of $R_{EE}$ (left) and $\Delta \lambda$ (right) for different network sizes (dots: raw data from simulations, line: mean, shading: standard deviation). Above a certain clustering threshold, SSA emerges and increases as  $R_{EE}$ grows; the intensity of the SSA dynamics is in line with the presence of an eigenvalue gap $\Delta \lambda$ in the weight matrix.
\textbf{B} Spike-rate variability as a function of $R_{EE}$ (left) and $\Delta \lambda$ (right) for a network of 2000 neurons with different numbers of clusters, yielding qualitatively similar results (dots: raw data from simulations, line: mean, shading: standard deviation).
\textbf{C} Relationship between the clustering strength $R_{EE}$ and the spectral gap $\Delta \lambda$. Observe that $R_{EE}$ is not sufficient to determine $\Delta \lambda$, i.e. $\Delta \lambda$ is influenced by other aspects such as the network size and number of groups.}
  \label{fig:2}
\end{figure}

In contrast, the LIF network with clustered excitatory neurons displays banded SSA dynamics. Spectrally, its weight matrix exhibits a clear gap $\Delta \lambda$ along the real axis of its spectrum (Figure \ref{fig:1}B, right) and, as shown below in Section~\ref{sec:Schur} (Fig.~\ref{fig:3_p3}), the associated Schur vectors also exhibit a measure of structural block-localization.
We remark that, in general, this should not be expected \textit{a priori}, since the weight matrices are \textit{asymmetric} and include both positive (excitatory) and negative (inhibitory) couplings.

To ascertain the dynamical relevance of the spectral gap, we examined the relation between the clustering strength in the network, defined as the ratio of probabilities of connections inside and outside the neural assemblies ($R_{EE} = p_\text{in}^{EE} / p_\text{out}^{EE}$), the spectral gap ($\Delta \lambda$), and the magnitude of the numerically observed SSA dynamics. To quantify the assembly spike-rate variability, we have defined two complementary metrics. 
First, the metric $\widehat S$ measures the heterogeneity in the firing rates of the putative cell-assemblies averaged over the simulation (see Eq.~\eqref{eq:S_hat} in Materials and Methods). A large value of $\widehat S$ indicates that the average firing rates of the assemblies are diverse, whereas a low $\widehat S$ indicates that all groups have very similar firing rates at all times and no group shows elevated firing.
Under SSA dynamics, the variability of firing rates across groups increases in time as the assemblies transition between high and low firing rates.
As discussed below in Section~\ref{sec:non-switching}, it is possible that the heightened firing activity is localized in a particular assembly and does not switch between assemblies.
This non-switching dynamics where only a group of neurons exhibits elevated firing over the entire simulation time can be detected using the second variability measure $\widehat S_T$, defined in Eq.~\eqref{eq:S_That}.
For SSA dynamics, both $\widehat S$ and $\widehat S_T$ give similar results (see Fig.~S1). 
In the rest of the paper, we focus on SSA dynamics and mainly use $\widehat S$, except when discussing the transition to non-switching behavior in Section~\ref{sec:non-switching}. 
Definitions of these quantities are given in Materials and Methods.

Figure~\ref{fig:2} shows that SSA becomes observable above a threshold of the clustering strength $R_{EE}$, although the dependence of this threshold on the size of the network and number of clusters does not seem to follow an obvious pattern.  
On the other hand, our numerics show that the spectral gap provides a more direct indicator of the presence of SSA dynamics in the network.
The relationship between $R_{EE}$ and $\Delta \lambda$ (Figure \ref{fig:2}C) shows that knowing $R_{EE}$ is not sufficient to determine $\Delta \lambda$ as the spectral gap depends on other factors such as the network size and the number of groups in the network.
Together with the examination of the structure of the associated orthogonal subspace (see Section~\ref{sec:Schur}), such spectral characterization may be used to establish the potential of networks to sustain SSA activity, even if there is no obviously clustered network model known \textit{a priori}.

The observed spectral properties of clustered LIF networks lead us to consider a stylized linear rate model for neuronal activity in the next section, as a means to gain insights into the defining factors of the wiring diagram leading to SSA.

\begin{figure}[tb!]
  \centering
  \includegraphics{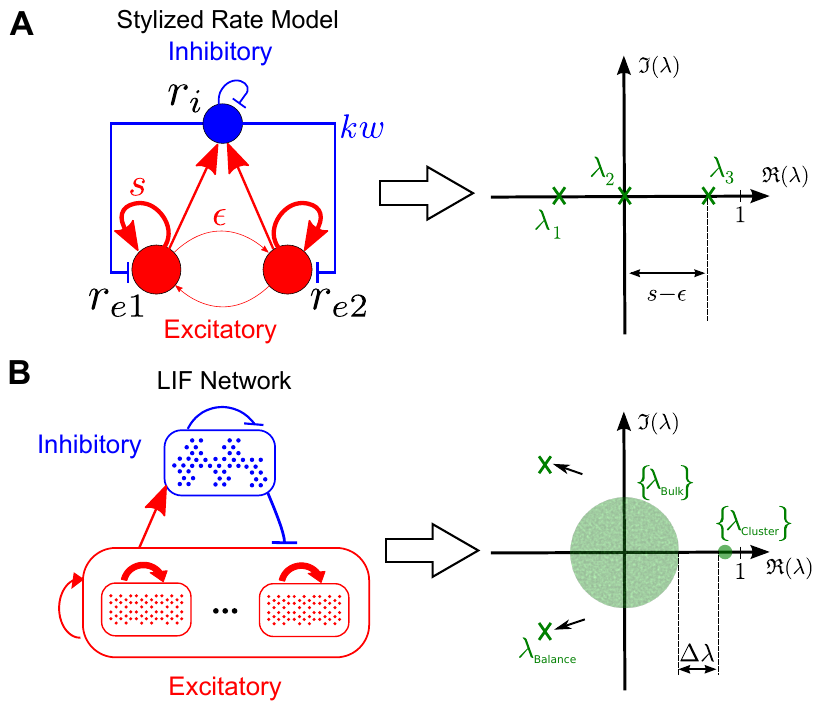}
  \caption{Structure of the eigenspectra of stylized and realistic networks that can exhibit SSA activity.
  \textbf{A} Schematic of the stylized model of a clustered network~\eqref{eq:rate_model}--\eqref{eq:W} 
  and illustration of the corresponding eigenvalues~\eqref{eq:eigen_Q}. 
  \textbf{B} Schematic of a more realistic clustered LIF network with the same wiring scheme on many groups and multiple neurons per group with an illustration of the associated spectrum.}
  \label{fig:3_p1}
\end{figure}

\subsection{Linear rate models and slow localized activity in clustered LIF networks}
\label{sec:linear_rates}
\subsubsection{A stylized linear rate model for networks with clustered excitatory neurons}
Motivated by the above findings, we proceeded to investigate how much of the observed LIF dynamics can be described by simple rate models.
From above, we expect that the spectral properties of the network lead to SSA dynamics that is coherent over, and transitioning between, clusters. 
Hence we consider simple rate models on coarse-grained networks where each node describes the behavior of a cluster.
As will become clear below, this reduced description allows us to identify the mechanism underpinning the onset of SSA activity without having to consider a plethora of statistical and technical detail. 
Furthermore, our main conclusions can be easily extended to multiple groups of clustered excitatory neurons, or to scenarios with probabilistic connectivity between the nodes (see e.g. Ref.\cite{Murphy2009}).

To explore this idea in the simplest setting, consider a linear firing rate model with three groups: two groups of clustered excitatory neurons and a group of inhibitory neurons with uniform coupling. This wiring scenario is abstracted in the form of a three-node network (Figure \ref{fig:3_p1}A), where each node represents a group of neurons, and a simple rate model is given by the following equations:
\begin{equation}
\label{eq:rate_model}
  \tau \dfrac{d\mathbf{r}}{dt} = -\mathbf{r} + W\mathbf{r} + \boldsymbol{\xi} = -(I-W)\mathbf{r}+\boldsymbol{\xi}.
\end{equation}
Here $\mathbf{r} = (r_{e1}, r_{e2}, r_{i})^T$ are the firing rates of the neuron groups; $W$ is the synaptic connectivity weight matrix; and $\boldsymbol{\xi}$ is a random input to the network.
The firing rates are measured relative to some baseline activity, and may be positive or negative.

To describe this topology with two excitatory clusters we define a 
connectivity matrix $W$ of the form:
\begin{gather}
\label{eq:W}
W = \begin{pmatrix} s & \epsilon & - kw\\ \epsilon & s & - kw\\ w/2 & w/2 & -kw \end{pmatrix}, \\
\label{eq:ineq_W}
s>\epsilon>0, \quad k\geq 1, \quad w=(s+\epsilon).
\end{gather}
The first inequality in~\eqref{eq:ineq_W} guarantees that the \textit{coupling strength} is positive:
$s-\epsilon>0$, i.e., the cross-coupling between the excitatory groups is weaker than the intra-group coupling, as would be expected from the notion of a cluster.
The second inequality in~\eqref{eq:ineq_W} ensures a balanced network (while keeping $k\approx 1$),  such that each group has at least the same amount of inhibitory and excitatory input: 
$\sum_j W_{ij} = (1-k)w \leq 0, \, \forall i$. 

To understand the dynamical behavior of this system, we evaluate the spectral properties of the connectivity matrix $W$~\cite{Galan2008,Murphy2009,Goldman2009}.
Note that we are dealing with a non-normal system, in which the eigenvectors may not be orthogonal and thus will provide a possibly misleading description of the dynamics~\cite{Trefethen2005}.
We therefore consider the Schur decomposition of $W$, which yields an orthonormal basis of the system~\cite{Murphy2009,Goldman2009}, i.e., it finds an orthogonal matrix $U$ and an upper triangular matrix $Q$ such that $W = UQU^T$.
Here $U$ contains an orthonormal basis while $Q$ contains the eigenvalues of $W$ on the diagonal and non-zero elements above the diagonal accounting for effective feedforward connections between the different modes \cite{Murphy2009,Goldman2009}.

For the matrix~\eqref{eq:W}, 
the associated upper triangular Schur form $Q$ is: 
\begin{gather}
\label{eq:Q}
Q = \begin{pmatrix} -w^+ & w_{\text{\textit{ff}}} & 0\\  & 0 & 0\\ & & s-\epsilon \end{pmatrix}, \\
w^+ = w(k-1), \quad w_{\text{\textit{ff}}} = (k+1/2)(w+\epsilon),  \nonumber
\end{gather}
and the matrix containing the orthonormal (Schur) basis is 
$U = \begin{pmatrix} \mathbf{u}_1 \,  \mathbf{u}_2  \, \mathbf{u}_3 \end{pmatrix}$ given by
\begin{equation}
\mathbf{u_1} =\dfrac{1}{\sqrt{3}} \begin{pmatrix} 1\\ 1\\ 1 \end{pmatrix},  \, 
\mathbf{u_2}= \sqrt{\dfrac{2}{3}}\begin{pmatrix}0.5\\ 0.5\\ -1 \end{pmatrix}, \, \mathbf{u_3} = \dfrac{1}{\sqrt{2}} \begin{pmatrix} -1\\ 1\\ 0 \end{pmatrix}.
\end{equation}
Each of the Schur vectors $\mathbf{u_i}$ represents a pattern of rates, a particular mode of firing on the network. The first mode $\mathbf{u_1}$ is associated with an overall mean firing pattern, while the second mode $\mathbf{u_2}$ corresponds to a relative difference in firing between inhibitory and excitatory groups. Note that in both cases the firing patterns of the two groups of excitatory neurons are aligned.  In contrast, the third mode
$\mathbf{u}_3$ describes an antagonistic activity localized on the excitatory neuron groups 
(i.e., when the firing rate of one excitatory group increases the other decreases, 
and vice versa) with the inhibitory node remaining unaffected at a baseline firing rate,
precisely in line with the SSA behavior observed in the 
full LIF network (Fig.~\ref{fig:1}A).
Observe also that, while the first two modes are coupled via an off-diagonal term in $Q$ (i.e., an effective feedforward connection \cite{Murphy2009,Goldman2009}), the switching behavior described by the third mode is uncoupled from the other system modes.

The structure of the Schur decomposition provides us with insight into the dynamics of the
model~\eqref{eq:rate_model}.
From~\eqref{eq:Q}, the eigenvalues of $W$ are:
\begin{equation}
\label{eq:eigen_Q}
  \lambda_1 = -w^+  \leq \lambda_2 = 0 < \lambda_3 = s-\epsilon,
\end{equation}
whence the eigenvalues governing~\eqref{eq:rate_model} are $\eta_1 = -1-w^+$, $\eta_2 = -1$ and $\eta_3 = -1+(s-\epsilon)$. For the system to be linearly stable, we require $\eta_3 < 0$, which implies that
the clustering strength in a stable system is constrained to be
\begin{equation}
\label{eq:strength}
0 < s - \epsilon < 1.
\end{equation}
Without an external input, the three modes decay exponentially with time constants $\tau_i = -1/\eta_i = 1/(1-\lambda_i)$. Therefore, it is possible to slow down the time scale associated with the SSA mode $\mathbf{u}_3$ by increasing the clustering strength $s-\epsilon \to 1$.  The perturbations introduced  by the random input induce mode-mixing and, in particular, break the symmetry between the firing rates of the two groups of excitatory neurons.  These perturbations excite the SSA mode $\mathbf{u_3}$, which will decay with time scale  $1/(1-\lambda_3)$ towards the solution with uniform rates on all groups ($\mathbf{r}^* = 0$).

In conclusion, the stronger the clustering (i.e., the larger $s-\epsilon< 1$), the slower the decrease of the asymmetric transients, and the more prominent SSA dynamics becomes. Note that this slow time scale in the activity patterns is directly associated with the eigengap $\lambda_3-\lambda_2=s-\epsilon$ (Fig.~\ref{fig:3_p1}), which is dictated by the clustering strength, in agreement with our observations relating the presence of SSA with $\Delta \lambda$ in Figure~\ref{fig:2}. The insights gained from this simple model have testable implications for LIF networks and lead to further ideas for neuro-physiological network wiring structures, which we explore in the following sections.

\begin{figure}[tb!]
  \centering
  \includegraphics{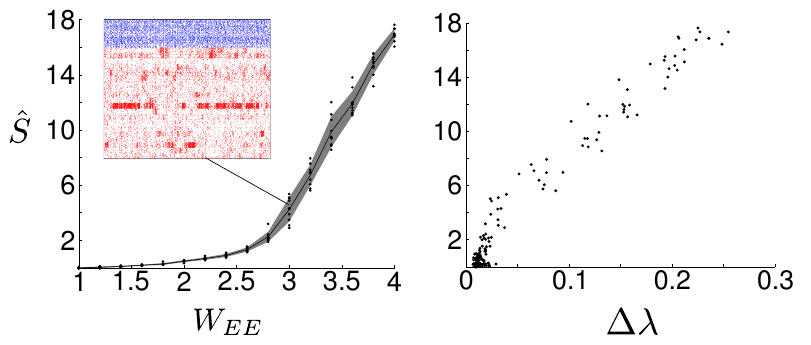}
  \caption{SSA dynamics can be achieved by changing solely the synaptic strengths even when the topological connections are kept uniform. 
  To create clustered networks, the ratio $W_{EE} = w_\text{in}^{EE} / w_\text{out}^{EE}$ was varied, where $w_\text{in}^{EE}$ and $w_\text{out}^{EE}$ refer to the in-group and out-group synaptic weights, respectively.
The spike-rate variability $\widehat S$ measuring the intensity of SSA dynamics is shown as a function of $W_{EE} $ (left - dots: raw data from simulations, line: mean, shading: standard deviation) and the spectral gap $\Delta \lambda$ (right).
Note that the connection probability is uniform for all the simulations, \textit{i.e.} $R_{EE}=1$, and
only the clustering of the weights is varied. }
  \label{fig:3_p2}
\end{figure}

\subsubsection{The linear rate model and the full dynamics of clustered LIF networks}
\label{sec:weightsMechanism}

Our rate model can be extended to networks with $c$ groups and 
the overall structure of the eigenspectrum of $W$ will essentially retain its features (Figure \ref{fig:3_p1}; see \cite{Murphy2009} for a related discussion). 
First, there will be a `negative' eigenvalue ($\lambda_1$ in~\eqref{eq:eigen_Q} related to the pair $\lambda_{\text{Balance}}$ in Fig.~\ref{fig:3_p1}B), which reflects the fact that the network is balanced and stable, \textit{i.e.} the associated global activity mode decays.
Second, there will be a bulk of `small' eigenvalues centered around the origin ($\lambda_2$ in~\eqref{eq:eigen_Q} and in general the set $\{\lambda_\text{Bulk}\}$ in Fig.~\ref{fig:3_p1}B), which stem from the random connectivity present in the network---a random matrix has a circular eigenvalue distribution around the origin).
Finally, there will be an eigenvalue gap separating a small set of $c-1$ `large positive' eigenvalues ($\lambda_3$ in~\eqref{eq:eigen_Q} and $\{\lambda_\text{Cluster}\}$ in Fig.~\ref{fig:3_p1}B) from the bulk of the spectrum. 
These are the `slow' eigenvalues associated with an orthogonal Schur subspace with a block structure coherent with the clusters, and are thus responsible for the appearance of a new time scale in the system originating the observed SSA dynamics. This structure of the eigenvalues can be seen in the spectrum of the LIF network in~Figure~\ref{fig:1}B.

Inspired by the analysis of the linear rate model, we investigate how the results translate 
to the full nonlinear LIF dynamics.
First, it is clear from above that only the difference between the effective intra- and inter-group coupling strengths is essential for the observed slow activity. Such difference can be the result of 
clustered topological connectivity as in Ref.~\cite{Litwin-Kumar2012}, but can be equally achieved keeping the topology uniformly connected and tuning the synaptic weights.
To assess this possibility, we conducted numerical simulations of LIF networks with uniform connection probabilities between excitatory neurons, yet with larger intra-group synaptic weights while keeping the average weight constant (see Materials and Methods). 
Figure \ref{fig:3_p2} shows that tuning the weights towards a clustered structure (as given by the weight ratio $W_{EE} = w_\text{in}^{EE} / w_\text{out}^{EE}$) also leads to the emergence of SSA dynamics linked to a spectral gap.
Although unsurprising from a linear systems perspective, this result confirms the applicability of the spectral characterization to nonlinear LIF networks and, specifically, establishes its relevance for SSA dynamics in networks where only synaptic weights can be tuned.

\begin{figure*}[tb!]
  \centering
  \includegraphics{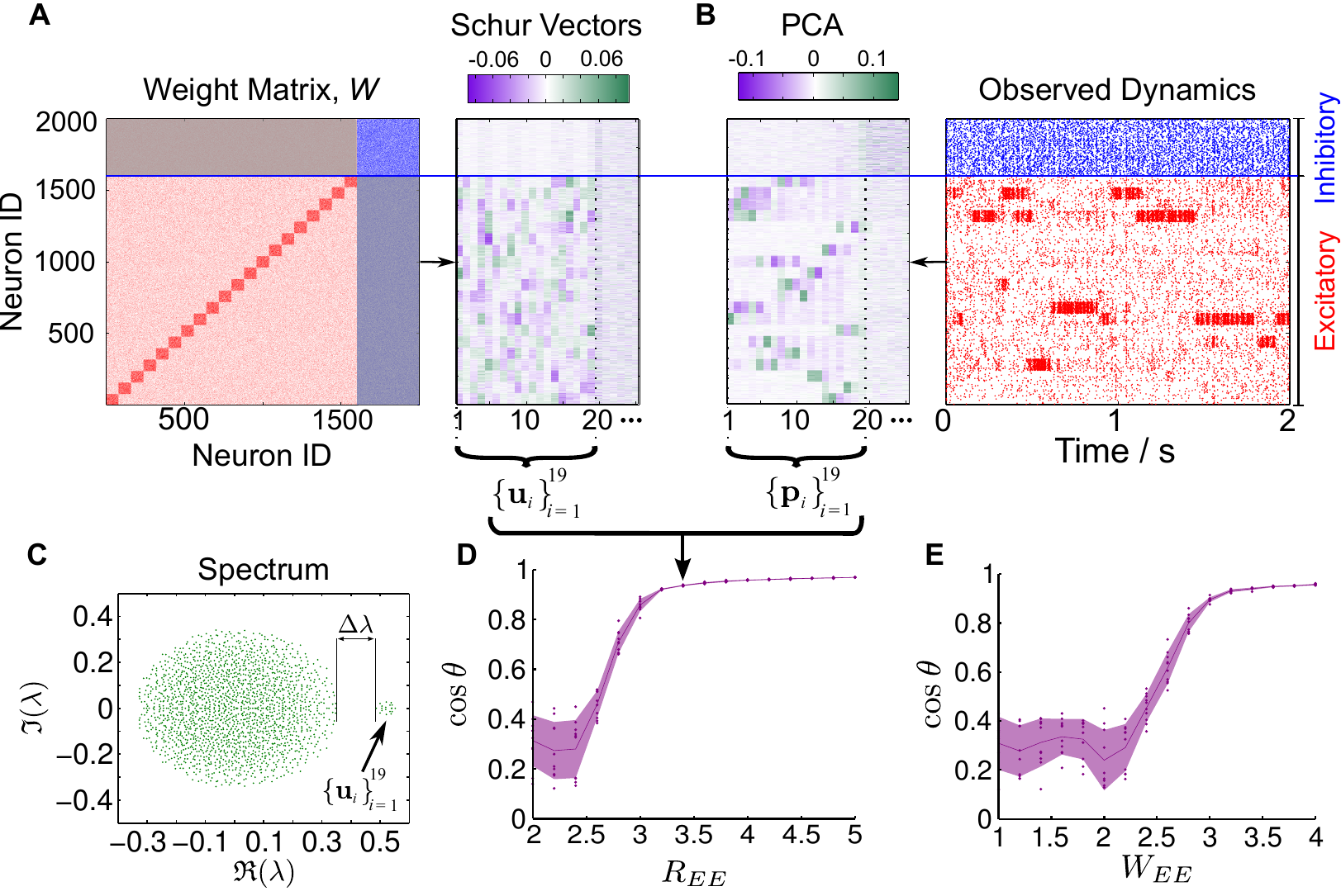}
  \caption{Unraveling the association between the leading Schur vectors of the weight matrix and the observed SSA activity.
 \textbf{A} Illustration of a weight matrix of the $N=2000$ neuron LIF network with $c=20$ groups of excitatory neurons ($R_{EE}=3.4$) shown together with the heatmap of the real parts of its first 25 Schur vectors. 
 The first 19 Schur vectors exhibit block-uniform patterns relatively constant within each of the 20 groups of neurons, whereas no such pattern is observed for the other Schur vectors. 
 Note that there is no pattern discernible over the inhibitory neurons.   
 The leading Schur vectors correspond to the cloud of `slow' eigenvalues above the gap, as indicated in \textbf{C}, and span a patterned dominant subspace that induces grouped dynamics in the network.
 \textbf{B} A long simulation (80s) of the LIF network dynamics (only the first 2s shown) was analyzed using PCA and the first 25 principal components (PCs) are shown.  Reflecting the banded structure of the simulated dynamics, the leading PCs also show a block-patterned structure consistent with the neuronal groups.
 \textbf{C}~On the spectrum of $W$, we indicate the group of leading eigenvalues above the gap associated with the dominant subspace.
 \textbf{D}   The alignment between the dominant Schur subspace of the $W$ matrix and the subspace of the strongest principal components is measured by the first principal angle $\theta$~\eqref{eq:theta}. 
 Above a threshold of the clustering strength $R_{EE}$, both subspaces become highly aligned in line with the observations in Fig.~\ref{fig:2} (dots: raw data from simulations; line: mean; shading: standard deviation).
\textbf{E} The same effect is observed when the clustering is introduced in the weights by varying $W_{EE}$, as in Fig.~\ref{fig:3_p2}. }
\label{fig:3_p3}
\end{figure*}

\subsubsection{The block-localization of the dominant linear subspace}
\label{sec:Schur}

We now consider a critical factor in the emergence of SSA, namely, that the leading Schur vectors of the weight matrix of LIF networks exhibit a measure of consistent block-localization on the groups of neurons which then exhibit cell assembly dynamics (Figure \ref{fig:3_p3}). 
Due to the non-normality outlined above \cite{Murphy2009,Goldman2009}, we consider orthogonal Schur vectors and not eigenvectors.
A typical Schur vector corresponding to one of the eigenvalues in the leading group $\{\lambda_\text{Cluster}\}$ exhibits a coherent pattern on each of the groups of the network, such that all neurons within a particular group are driven uniformly. This block-uniformity results in the grouped dynamics that characterizes SSA.  Representative patterns are shown in Figure~\ref{fig:3_p3}A.
In contrast, the Schur vectors from the bulk of the spectrum $\{\lambda_\text{Bulk}\}$ show no specific pattern of localization on any group of neurons. 
Note that the Schur vectors show no localization pattern on the inhibitory neurons either, in line with our analysis. 
The block-localization of the leading Schur vectors constitutes an additional spectral characterization to assess the emergence of SSA dynamics from the weight matrix alone. 

In Figure~\ref{fig:3_p3}, we quantify how the patterns observed in the SSA dynamics align with the dominant Schur vectors of the weight matrix. To do this, we first performed a principal component analysis (PCA) of the simulated firing rates
of the full LIF network and extracted the top activation patterns $\{\mathbf{p_1},\ldots,\mathbf{p_c}\}$ corresponding to the first principal components (see Figure~\ref{fig:3_p3}B and Materials and Methods).
To measure the alignment of the data patterns $\{\mathbf{p_i}\}$ with the dominant Schur vectors $\{\mathbf{u_1},\ldots,\mathbf{u_c}\}$ of the weight matrix $W$ corresponding to the leading eigenvalues above the gap, we computed the \textit{first principal angle} $\theta$ between these two subspaces: the smaller the angle, the closer the alignment \cite{Golub1996,Stewart2001}. 
Figure \ref{fig:3_p3}D shows that the dominant subspace of the Schur vectors of the weight matrix ($W$) are indeed aligned with the observed patterns and consistent with 
the embedded groups of neurons in the network. 
Finally, Figure \ref{fig:3_p3}E shows again that tuning the synaptic weights to cluster the network while keeping the topology uniformly connected renders a similar outcome.

\begin{figure}[htb!]
 \centering
 \includegraphics{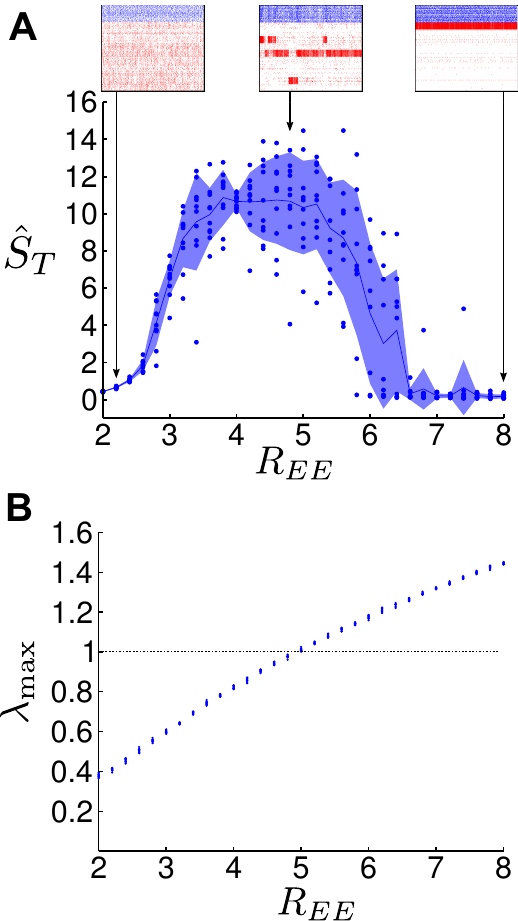}
 \caption{Effect of large clustering strength on SSA activity.
 \textbf{A}~ Large values of the clustering $R_{EE}$ lead to linear instability of the SSA dynamics and localization of the activity on one assembly. As measured by the spike rate variability across time ($\hat S_T$), the increase of $R_{EE}$ leads to SSA (signalled by the increased value of $\hat S_T$).  If the clustering increases further, $\hat S_T$ decreases, as the dynamics becomes dominated by one assembly only. The dots denote raw data from simulations (line: mean; shading: standard deviation). Inset: examples of raster plots for three data points in the three regimes.  The analysis corresponds to a clustered LIF network of 1000 neurons.
 \textbf{B}~ Plot of the eigenvalue with the largest real component $\lambda_\text{max}$ against the clustering strength $R_{EE}$. The linear condition $\lambda_\text{max} > 1$ is a good indicator of the dynamics becoming dominated by one cell assembly.}
 \label{fig:S2}
\end{figure}

\subsubsection{Increasing the clustering beyond the linearly stable regime: one dominant assembly}
\label{sec:non-switching}
Until now, our linear analysis has concentrated on the relevant regime for SSA, where the system is linearly stable~\eqref{eq:strength}: $\lambda_3 = s - \epsilon < 1.$ 
However, it is possible to increase the clustering strength ($s-\epsilon$), so that the intra-cluster connections are much stronger than the inter-cluster connections, and the linear system has an unstable eigenvalue. For the LIF network, if we increase $R_{EE}$ or $W_{EE}$ so that $\lambda_\text{max} \in \{\lambda_\text{Cluster}\} \geq 1$, then the associated localized firing mode, once excited, will activate itself. This regime can thus lead to a `winner takes all' solution, where one cell assembly dominates the firing of the whole network and suppresses all other neurons \cite{Rutishauser2011}.  

From the perspective of the observed dynamics of the LIF network, an increase of the clustering strength ($R_{EE}$ or $W_{EE}$) initially leads to the development of the eigengap for $\{\lambda_\text{Cluster}\}$ and, 
the emergence of SSA dynamics.  However, when the clustering strength becomes too large (and the largest eigenvalue goes beyond 1), a single assembly starts to dominate the firing pattern and the firing variability is reduced across time, an effect which is captured by a strong reduction in $\widehat S_T$, as illustrated in Figure \ref{fig:S2}. 
This behavior, which has been observed previously~\cite{Litwin-Kumar2012}, is thus also closely linked to the spectral properties of the system.
Note that the linear condition $\lambda_\text{max} \geq 1$ is only indicative: the non-linearity and boundedness of LIF dynamics can lead to (non-linearly stable) SSA dynamics beyond this condition.

\subsection{Beyond clustered excitatory neurons: SSA dynamics involving both excitatory and inhibitory neurons.}
\label{sec:alt_mech}
So far we showed that a linear rate model can provide valuable insights into the mechanisms underpinning SSA dynamics in networks with clustered excitatory neurons.
The crucial point for the emergence of SSA is the separation of time scales, dictated by the splitting of the leading eigenvalues of the weight matrix, together with the block-localization of the associated dominant subspace. 
These spectral properties can be introduced into LIF networks by entirely different synaptic couplings, and 
we now consider a mechanism in which the inhibitory neurons are involved more centrally in generating SSA dynamics.

\begin{figure}[tb!]
  \centering
  \includegraphics{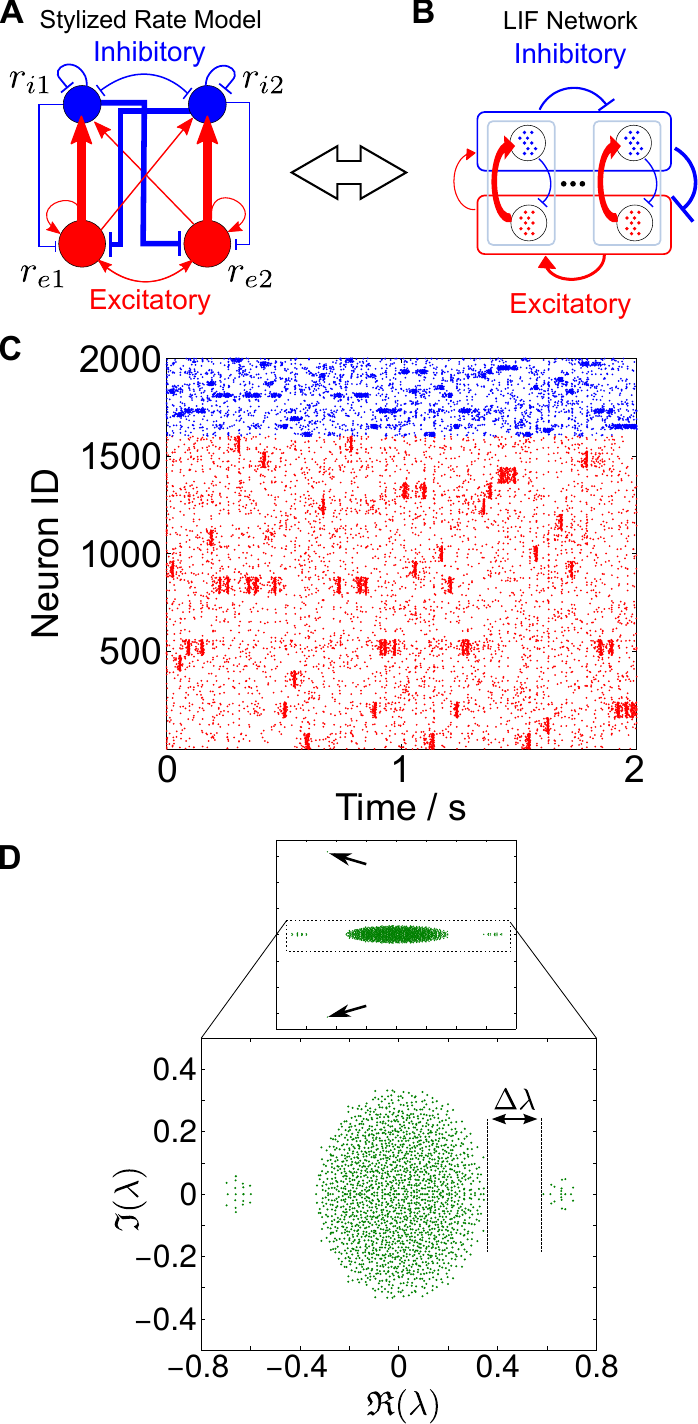}
  \caption{
  SSA dynamics can result from excitatory-to-inhibitory feedback loops.
  \textbf{A}~Schematic of a stylized network model with two excitatory-to-inhibitory feedback loops corresponding to the model~\eqref{eq:Winh}.
  \textbf{B}~Illustration of the corresponding LIF network with such a co-clustered feedback mechanism between neuron types.
  \textbf{C}~Raster plot of the dynamics of a co-clustered LIF network with $N=2000$ neurons and $c=20$ pairs with excitatory-to-inhibitory feedback. Note that the inhibitory neurons also exhibit SSA dynamics here.
  \textbf{D} Spectrum of the weight matrix exhibiting a spectral gap both on the left and right hand side of the bulk of the spectrum (see text for details).}
  \label{fig:4}
\end{figure}

Consider the wiring diagrams presented in Figure \ref{fig:4}A-B, representing the simple rate model and its corresponding LIF schematic. In this case, the wiring does not involve clustered coupling between groups of excitatory neurons, but rather relies on preferential connectivity patterns \textit{between} inhibitory and excitatory neuron groups only. Each group of excitatory neurons activates preferentially an associated group of inhibitory neurons and, in turn, this group of inhibitory neurons feeds back more weakly to its associated group of excitatory neurons (see Materials and Methods for a full description of the network).  
Therefore the effective functional circuitry consists of both excitatory and inhibitory neurons embedded in a feedback loop and, as a consequence, the inhibitory neurons play an integral role in generating the spatio-temporal dynamics and display SSA dynamics too, as we show below.

This wiring mechanism was suggested by the stylized firing rate model with two coupled pairs of inhibitory/excitatory feedback loops in Figure \ref{fig:4}A.  This system is described by Eq.~\eqref{eq:rate_model} with a vector of firing rates for the four groups $\mathbf{r} = (r_{e1}, r_{e2}, r_{i1}, r_{i2})^T$ and a synaptic weight matrix defined as:
\begin{gather}
\label{eq:Winh}
W=  \begin{pmatrix} w & w & - k\epsilon & -ks\\ w & w & -ks & -k\epsilon\\ s & \epsilon & -kw & -kw\\ \epsilon & s & -kw & -kw \end{pmatrix}, \\
s > \epsilon, \quad k\geq 1, \quad w = (s+\epsilon)/2. 
\end{gather}
Similarly to~\eqref{eq:W}, the system is balanced and $s - \epsilon$ captures the clustering strength within the excitatory-inhibitory pairs.
The Schur decomposition of $W$ leads to the following Schur form
\begin{gather}
Q = \begin{pmatrix} -w^+ & w_{\text{\textit{ff}}} & &\\  & 0 & &\\ & &\sqrt{k} (s-\epsilon) & w_{\text{\textit{ff},2}} \\ & & &-\sqrt{k} (s-\epsilon) \end{pmatrix}, \\
w^+ = (k-1)(s+\epsilon), \nonumber \\ 
\quad w_{\text{\textit{ff}}}=(k+1)(s+\epsilon),  \quad w_{\text{\textit{ff},2}} = -(k-1)(s-\epsilon), \nonumber
\end{gather}
where the eigenvalues of $W$ are on the diagonal.
When the leak term is considered, the largest eigenvalue of~\eqref{eq:rate_model} 
is ($-1 + \sqrt{k} (s-\epsilon)$). Hence, to keep the system stable we need to constrain
\begin{equation}
 0 < \sqrt{k} \, (s-\epsilon) < 1,
 \end{equation}
 and the spectral gap is again controlled by $s - \epsilon$.
In the associated orthonormal Schur basis, the first two modes of the firing rate dynamics
\begin{equation}
    \mathbf{u_1} =\dfrac{1}{2}\begin{pmatrix} 1\\ 1\\ 1\\ 1 \end{pmatrix}, \quad \mathbf{u_2} = \dfrac{1}{2}\begin{pmatrix}1\\ 1\\ -1\\ -1 \end{pmatrix}
\end{equation}
are again global `sum' and `difference' modes, which interact via a balanced amplification mechanism~\cite{Murphy2009}.
However, there are also two localized modes $\mathbf{u}_3$ and $\mathbf{u}_4$ that can lead to slow structured activity:
\begin{equation}
\label{eq:modes_inhexc}
  \mathbf{u_3} =\dfrac{1}{\sqrt{2k+2}}\begin{pmatrix} \sqrt{k}\\-\sqrt{k}\\ 1\\ -1 \end{pmatrix}, \quad\mathbf{u_4} = \dfrac{1}{\sqrt{2k+2}}\begin{pmatrix}1\\ -1\\ -\sqrt{k}\\ \sqrt{k} \end{pmatrix}.
\end{equation}
Of these, $\mathbf{u_3}$ is associated with the largest eigenvalue and describes the slow(est) dynamics. This mode corresponds to a firing pattern of correlated activity within the pairs $(r_{e1},  r_{i1})$  and $(r_{e2}, r_{i2})$, and anti-correlated activity across the pairs.
As before, this analysis extends to networks with more groups and/or stochastic coupling between groups (for related arguments see supplementary material of Ref.~\cite{Murphy2009}).

\begin{figure}[tb!]
  \centering
  \includegraphics{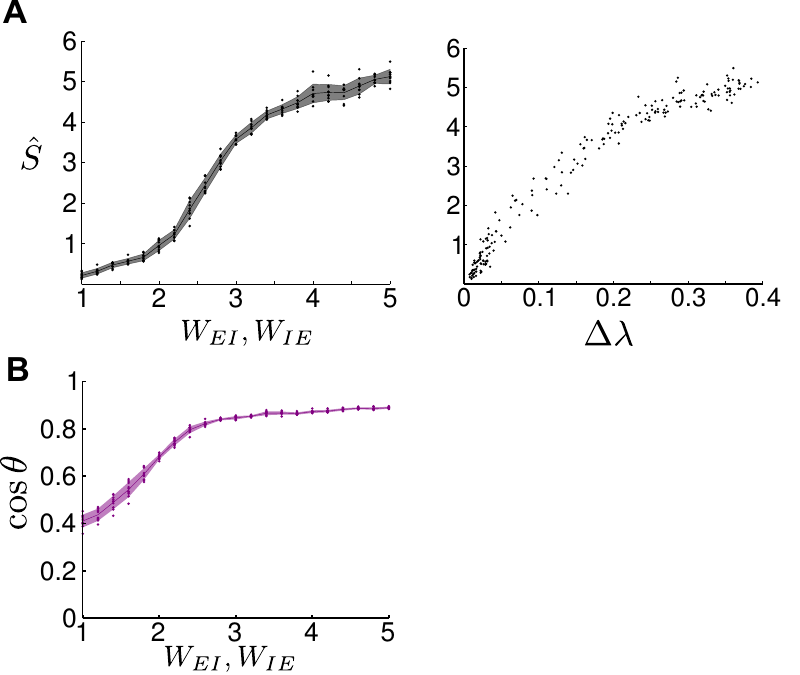}
\caption{ The presence of SSA dynamics in LIF networks with excitatory to inhibitory feedback loops.
  \textbf{A} The spike-rate variability, which measures the intensity of SSA dynamics, as a function of $W_{EI} = W_{IE}$ (left; dots: raw data from simulations, line: mean, shading: standard deviation) and $\Delta \lambda$ (right). See text and Fig.~\ref{fig:3_p2} for further details.
  \textbf{B} The first principal angle between the subspaces of the principal firing patterns and the dominant Schur vectors of the weight matrix $W$ show high alignment. See text and Fig.~\ref{fig:3_p3} for further details.
}
  \label{fig:4_part2}
\end{figure}

In Figure \ref{fig:4}C-D, we show that the implementation of this wiring mechanism into a full LIF network displays SSA dynamics with a spectral gap, as predicted by our simple rate model, and the paired excitatory/inhibitory neurons act as a functional circuit displaying synchronous firing behavior. 
Importantly, note that, in contrast to the excitatory clustered scenario, there are no groups with dense reciprocal couplings in this network topology.  
Our numerics also show that varying the strength of the functional grouping ($W_{EI} = W_{IE}$) leads to the emergence of SSA dynamics, linked to the presence of the spectral gap $\Delta \lambda$ (Fig.\ref{fig:4_part2}A), and to the alignment of the leading Schur vectors with the `correct' functional circuits, i.e., pairs of excitatory and inhibitory groups together (Figure \ref{fig:4_part2}B).

It is interesting to note that in this topology there is also a group of eigenvalues bounded away in the negative direction (see Figure~\ref{fig:4}D). These modes are the quickest decaying and correspond to `anti-correlated' firing states, in which excitatory neurons act in synchrony with the inhibitory groups to which they are not functionally associated. Such fast decay reinforces the survival of synchrony within the functional groups in the network.
Interestingly, these modes were already present in our linear rate model: $ \mathbf{u_4}$ in~\eqref{eq:modes_inhexc} shows the same firing pattern with the fastest eigenvalue $\lambda_4=-\sqrt{k} (s-\epsilon)$.

\subsection{SSA dynamics in LIF networks with alternative connectivity topologies}

\begin{figure}[tb!]
  \centering
  \includegraphics{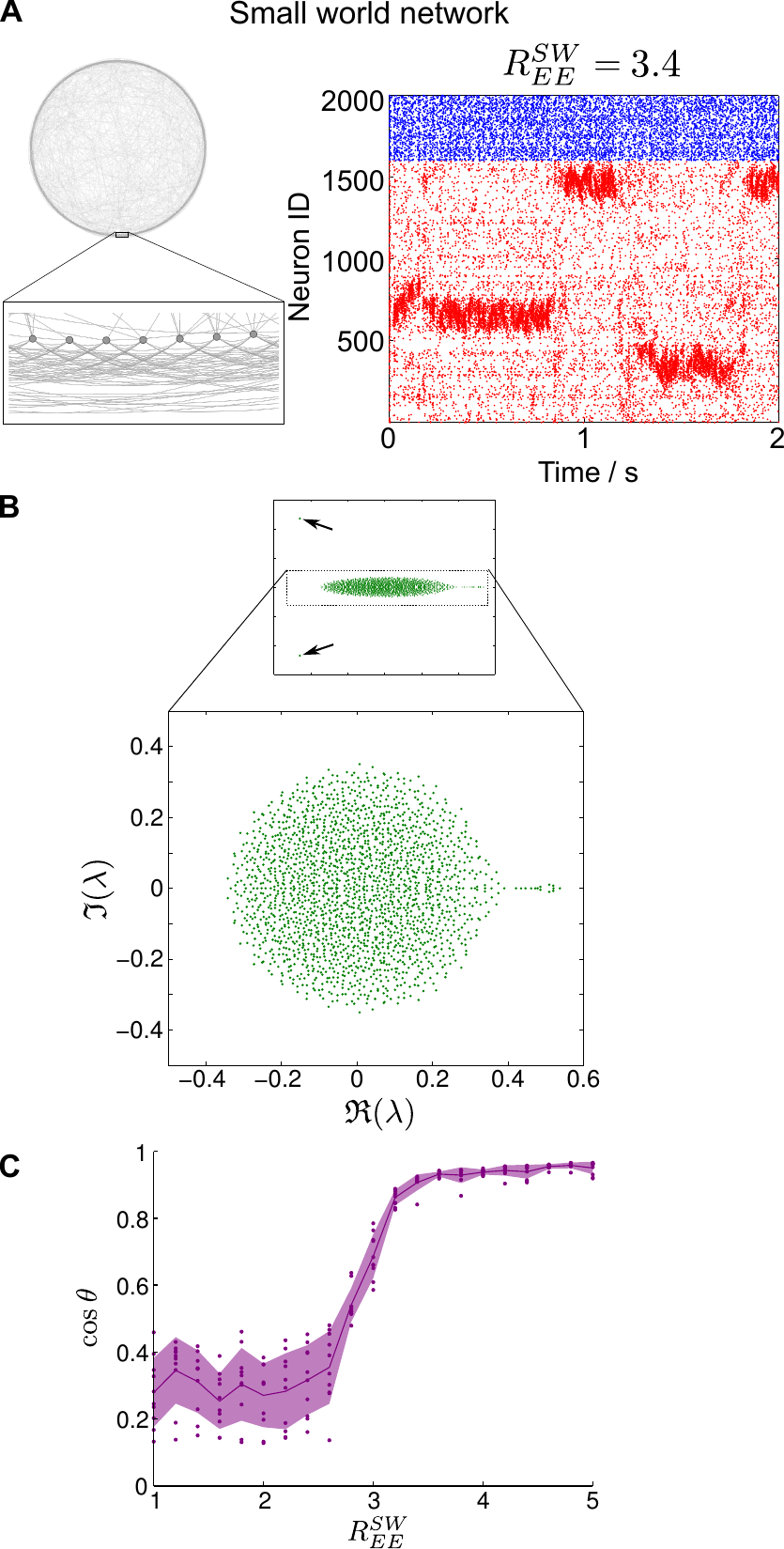}
  \caption{Small-world networks can exhibit SSA activity. 
  \textbf{A}~Small-world network and the resulting raster plot of its LIF network simulation.
  SSA activity is observed, yet with less distinct boundaries when compared to the clustered case.
  \textbf{B}~Spectrum of the weight matrix of the small-world network showing a small eigenvalue gap.
  \textbf{C}~The principal angle between the dominant subspace spanned by principal components of the firing rates and the dominant Schur vectors of the weight matrix shows that, as the modularity of the small-world rewiring becomes larger, there is alignment between the observed dynamics and the slow directions of the weight matrix  (dots: raw data from simulations; line: mean; shading: standard deviation).}
  \label{fig:5_SW}
\end{figure}

Although perhaps the most intuitive way of generating SSA dynamics follows from clustering the connectivity of the LIF network,  our analysis above suggests that alternative types of structural organization support SSA dynamics in LIF networks, as long as they are characterized by a gap in their eigenvalue spectrum and a measure of block-localization of the Schur vectors.
More generally, one could conjecture that any low rank perturbation of the weight matrix of a balanced network which leads to an eigenvalue gap might be a valid candidate for a mechanism to generate SSA in neuronal dynamics.
A few such network architectures are worth highlighting, as they have been considered of particular relevance in a neuroscience context.

\subsubsection{SSA dynamics in networks with small-world organization}

The first noteworthy example is the broad class of small-world like networks~\cite{Watts1998}, since small-world organizations have been observed in many structural and functional neurophysiological networks~\cite{Bullmore2009}.
While many modular networks can have the small-world property \cite{Meunier2010}, here we focus on the archetypal small-world structure \`{a} la Watts and Strogatz~\cite{Watts1998}, which does not display a distinct modular organization but instead may be seen as a mixture of a $k$-nearest neighbor ring lattice and a random graph.  
It is well known that such small-world networks possess distinctive spectral properties which have important implications for dynamical processes, such as synchronization~\cite{Barahona2002}.
As Figure \ref{fig:5_SW}A shows, small-world like LIF networks (see Materials and Methods for details of the construction) can indeed display SSA dynamics.
In line with our arguments above, the weight matrix has a leading group of eigenvalues separated from the bulk, which dictate the slow dynamics, although in this case the gap is not as large and the slow eigenvalues are not as tightly bunched (Figure \ref{fig:5_SW}B). 
These separated eigenvalues correspond to a subspace spanned by eigenmodes with a slow spatial variation along the `backbone' ring, and which lead to the localization and subsequent switching of the neural activity between subgroups of neurons.
The principal angle between the firing patterns and the dominant Schur vectors (Figure \ref{fig:5_SW}C) indicates that the firing patterns are again dictated by the spectral properties of the underlying small-world topology.
Note, however, that due to the lack of hard boundaries in the spectral groupings, both for the eigenvalue structure and the smoothly-varying dominant Schur vectors, 
the SSA dynamics (Figure \ref{fig:5_SW}A), although present, is not as distinctly localized as in the examples above.
This behavior is effectively a result of the heavy overlap induced by the underlying lattice-like pristine world in the small-world construction~\cite{Barahona2002}.

\subsubsection{Lack of SSA dynamics in scale-free networks}
Another class of networks that has attracted tremendous interest is so-called scale-free networks \cite{Barabasi1999}.
These networks are characterized by a fat-tailed degree distribution, a fact that has been observed empirically for a variety of networks.
As shown for undirected random graphs, the spectrum of scale-free networks may have a spectral gap, due to the presence of hubs with very large degrees \cite{Nadakuditi2013}.
However, the associated eigenvectors are usually localized around the hubs, and as such are not able to trigger a consistent pattern of localized activity across a group of nodes.
In our simulations of scale-free LIF networks, we did not observe SSA dynamics: the dynamics was essentially concentrated around the hub with no transitions in time  (\textit{i.e.}, only a core of neurons around the hub had consistently elevated firing rates), hence no SSA. Indeed, the networks had 
only one large eigenvalue separated from the bulk 
(due to the high degree of the main hub) and its associated Schur vector was highly localized around the hub \cite{Nadakuditi2013}.
An example of this hub-centric dynamics can be found in Figure S2 of the Supplementary Information.

\subsubsection{SSA dynamics with multiple time scales in LIF networks}

\begin{figure}[tb!]
  \centering
  \includegraphics{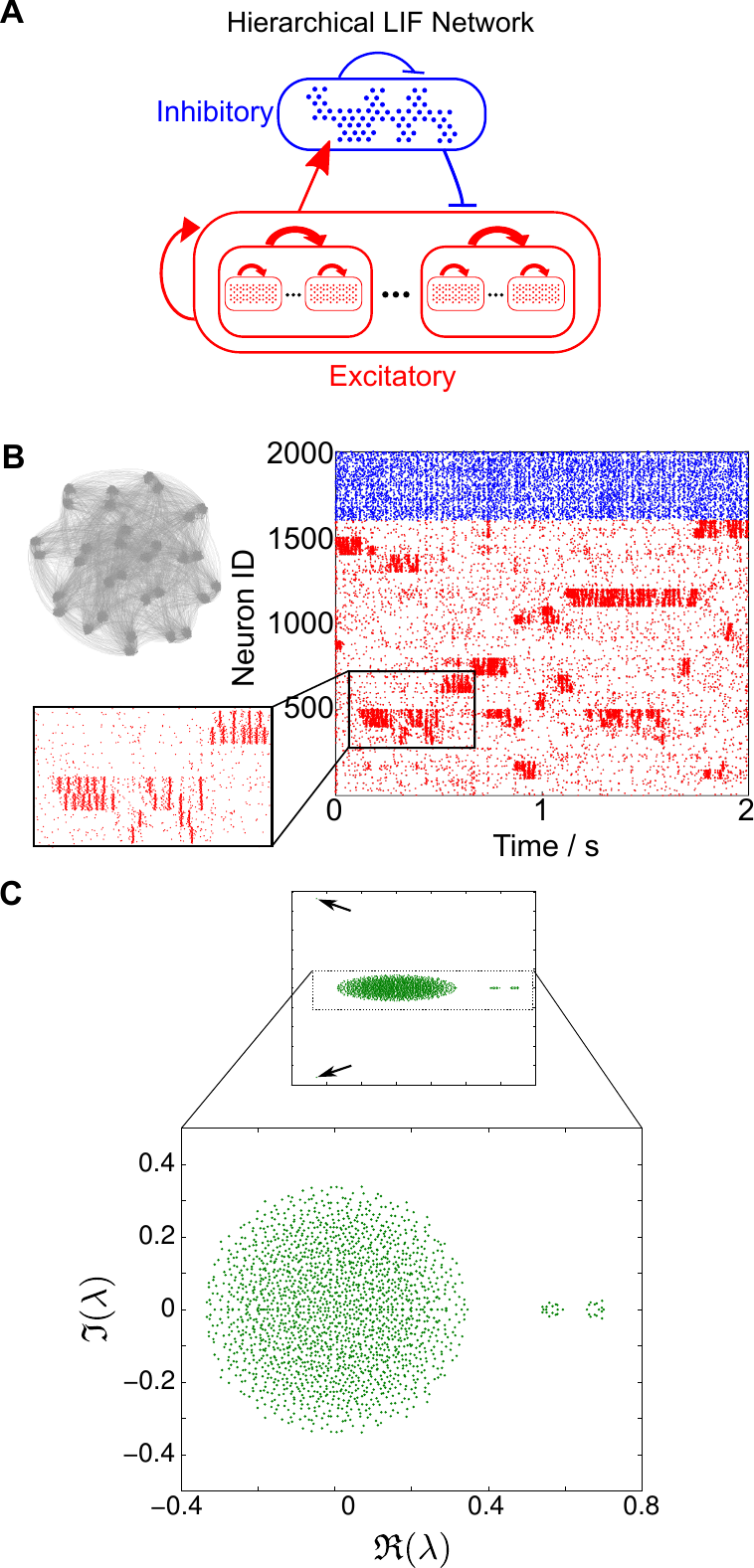}
  \caption{SSA activity in hierarchical LIF networks.
  \textbf{A}~Schematic of a hierarchically clustered LIF network. Each excitatory group consists of two subgroups.
  \textbf{B}~A hierarchical modular LIF network with $N=2000$ neurons and the resulting raster plot of its simulation. SSA activity is observed at two time scales, corresponding to the two hierarchical levels embedded in the network structure: slow switching between the large groups and faster switching between the inner subgroups (see inset).
  \textbf{C}~The spectrum of the weight matrix of this hierarchical network exhibits two eigenvalue gaps corresponding to the two slow time-scales the network can support.}
   \label{fig:5_hier}
\end{figure}

It is also possible to enforce hierarchical arrangements of the functional wiring mechanisms discussed so far, thus allowing for a variety of combinatorial arrangements.
Depending on the number of hierarchical layers, this enables the introduction of \textit{multiple time scales} in the spatio-temporal segregated activity. 
As an illustrative example, Figure \ref{fig:5_hier} shows the dynamics of a LIF network with a two-level hierarchy of clustered excitatory neurons. 
This network exhibits SSA with two `slow' time scales: a very slow switching between the groups of the top level of the hierarchy, and the not-so-slow pulsation of activity between the subgroups within each of the large groups, thus reflecting the second level of the hierarchy. 
We remark that one could have employed a combination of different `functional circuits' in the individual layers of the hierarchy leading to potentially different behaviors.
As real neural networks display multiple layers of hierarchical organization \cite{Shimono2014, McGinley2013,Savic2000,Ambrosingerson1990,Felleman1991}, understanding such schemes is of great importance for neural computation, and will be addressed in future work.

\section{Discussion}
Answering the question of how the wiring of neural circuits governs neural dynamics is a key step in understanding neuronal computations.
Here, we showed how knowledge of certain spectral features of  the matrix of neuronal connectivity can help understand what dynamics a network can support.

In this paper we focused on LIF networks exhibiting slow-switching assembly (SSA) dynamics, where groups of neurons show a sustained and coherent increase in their firing rates, with slow switching between epochs of localized firing in different groups across the network.
We found that the presence of the slow switching time scale is reflected in the spectral properties of the synaptic weight matrix: a gap separating the leading eigenvalues together with a block-localization of the associated Schur vectors on groups of neurons is a key indicator of the presence of SSA dynamics in the network. 
In line with this observation, multiple gaps in the eigenvalue spectrum are indicative of further time scales in the network dynamics (see Figure~\ref{fig:5_hier}). Moreover, when the leading eigenvalue becomes larger than one, the SSA dynamics becomes localized on one of the assemblies (see Figure~\ref{fig:S2}).

First, we revisited the case of balanced LIF networks with clustered excitatory neurons~\cite{Litwin-Kumar2012} and observed that only when there is an eigenvalue gap and Schur block-localization does the network display SSA dynamics.
Further analytical understanding from a stylized firing-rate model allowed us to determine that the clustering strength drives the development of the eigenvalue gap responsible for the slow switching between localized firing modes.
We remark that clustered excitatory connectivity leads to a Hebbian amplification regime~\cite{Murphy2009}: the more stable the pattern formation, the slower the dynamics of any banded activity. Fast switching between well-defined, stable patterns is thus only achievable for strong input changes. 

As suggested from our spectral characterization, and confirmed by simulations, we showed that SSA dynamics can be achieved through the structured tuning of synaptic strengths, rather than through the clustered rewiring of the connections.
While the equivalence between topological and weight organizations is clear for a linear system, such a direct correspondence is not guaranteed a priori for a non-linear system. 
In fact, the clustering of weights appears to have a slightly larger influence compared to topological organization in our simulations of non-linear LIF systems.
More importantly, this dynamical equivalence between clustered topology and clustered synaptic weights has potential ramifications for learning and synaptic plasticity; it indicates that the alteration of the weight structure can lead to the emergence of grouped activity without the need for structural rewiring of physical synaptic connections, thus suggesting a cost-effective adaptation to stochastically encode different firing patterns \cite{Laughlin2003}.

Within LIF networks with clustered excitatory connectivity~\cite{Litwin-Kumar2012}, inhibitory neurons only play a balancing background role, whereas experimental studies have revealed a vast diversity of interneuron subtypes that play key roles in computation~\cite{Kepecs2014,Roux2014}.
We thus investigated mechanisms in which inhibitory neurons have an active, functional role in generating SSA dynamics.
We demonstrated both analytically and via numerical simulations how such functional circuits can be constructed with the help of a feedback loop between excitatory and inhibitory neurons, so that  inhibitory neurons themselves exhibit SSA dynamics and become an integral functional part of the dynamics.
As fewer inhibitory neurons are present, adapting the weights according to this functional co-clustering scheme may also provide a cost effective alternative to generate SSA activity in neuronal networks.

The emergence of SSA dynamics can be achieved not only by clustering excitatory neurons but also by shaping the network structure in several ways.
For instance, a small-world network topology of the excitatory neurons is also able to support SSA dynamics due to the spectral properties of small-worlds, which induce a separation of the slow eigenvalues and the localization and switching of firing activity associated with slowly spatially-varying dominant Schur vectors.
In contrast, scale-free networks lack the block-localization of the Schur vectors on multiple groups which is necessary to consistently induce SSA dynamics in LIF networks.
Further topologies will be explored in the future. 
In particular, inhibitory neurons that inhibit other inhibitory neurons (so-called disinhibition patterns \cite{Silberberg2007,Xu2013,Lee2013,Pi2013,Pfeffer2013,Kepecs2014}) provide an interesting example. 
A different avenue might be provided by balanced amplification mechanisms \cite{Murphy2009}, which could be potentially used for creating grouped activity, yet without the introduction of a slow time scale.  Other interesting questions in this respect are:  which wiring mechanisms provide the most economical, or evolutionarily fit, variant~\cite{Niven2008, Laughlin2003} to induce a given dynamics; and how does the observed diversity of interneurons (and the multiple roles they can play) relate to this dynamical picture.

Our work emphasizes the importance of the spectral characterization of the weight matrix for the dynamics taking place on these topologies. Although spectral properties are also key in characterizing the dynamical response of networks operating at criticality~\cite{Larremore2011}, our observations here correspond to a different phenomenon. The emergence of a dominant assembly when $\lambda_\text{max} \gtrsim 1$ leads to a saturation of the network and a decrease of its dynamical heterogeneity, in contrast to systems of coupled non-modular excitatory units at criticality, which maximize their dynamic range for $\lambda_\text{max} =1$~\cite{Larremore2011,Beggs2003}.

Our work links up with experimental findings that cortical states can arise spontaneously and exhibit switching behavior. Recordings from anesthetized cat visual cortex have revealed activity states that dynamically switch, and these states match closely to recorded orientation maps \cite{Kenet2003}.
Such cortical states likely arise due to similarly tuned neurons having higher intra-cortical connectivity \cite{Ko2011,Harris2013}, as is also predicted through models of orientation maps \cite{BenYishai1995,Somers1995, Ernst2001}.
These observations are similar to our theoretical outcomes which indicate that dynamic transitioning between different network states can be driven and sustained based only on the underlying network topology.
Other experiments have identified states in CA3 network activity of the hippocampus \cite{Sasaki2007} where active cell assemblies exist for tens of seconds before sharply transitioning to a new state. However, these were metastable states (hence unlikely to be reactivated) and included a core population of cells consistently active in all states. 
While such complex dynamics are beyond the simple models presented here, it may be feasible to model such metastable dynamics through more elaborate network topologies, which may include synaptic plasticity~\cite{Litwin-Kumar2014}.
Finally, let us remark that our reduced model descriptions effectively focused on networks implementing a \textit{rate-based} coding mechanism and do not include spike-timing of cell assemblies, which has been identified to be an important component in neuronal computation, e.g. in rodents \cite{Harris2003,Foster2006}, monkeys \cite{Riehle1997}, songbirds \cite{Hahnloser2002}, and grasshoppers \cite{Rokem2006}. 
Whether our results can be translated to a time-coding regime would be an interesting question for future work. 

As connectomics continues to advance the mapping of connections and synaptic strengths in wide areas of the brain, experimentally obtained weight matrices can be analyzed spectrally as described above to determine if SSA dynamics can be supported by the networks under study.
While we focused here on the simplest mechanisms underlying SSA activity, future work could also consider the construction of biophysically realistic models that take into account the growing literature on the distribution of synaptic strengths, synaptic contacts, firing rates, and other relevant cortical parameters that tend to show a lognormal distribution~\cite{Song2005, Hromadka2008, Buzsaki2014, Oh2014}.

Although knowledge of the network structure (connectomics) is not sufficient to predict the dynamics a network circuit will display (for instance, the network dynamics can be dominated completely by a strong input), it can still give valuable insights on the firing patterns the network can support. 
Conversely, the observation of neuronal dynamics alone may not be sufficient to understand neural computation in detail, since different network topologies can yield similar dynamics.
Hence, our work hints at how connectomics and neuronal dynamics data can provide complementary and intertwined routes for systems neuroscientists to study the computational principles implemented by the brain. 

Finally, it is important to remark that in order to get a fuller picture of the relation between structure and dynamical network properties many other factors not considered here will be of interest, including the distribution of inputs, the precise location of synapses on the post-synaptic neuron, and the plasticity rules that govern the evolution of the network over time~\cite{Clopath2010}.
Linking the insight gained from our work to experimental observations or to highly detailed computational models \cite{Ahrens2013,Reimann2013} would thus be a fruitful next step in order to bridge the gap between neuronal structure, dynamics, and ultimately function.

\section{Material and Methods}
\subsection{Leaky integrate-and-fire networks}
We simulated leaky-integrate-and-fire (LIF) networks where the non-dimensionalized membrane potential of each neuron ($V_i(t), i=1,\dots,N$) was modeled by:
\begin{equation}\label{eq:LIF_model}
  \dfrac{d V_i(t)}{dt} = \dfrac{1}{\tau_m}(\mu_i - V_i(t)) + \sum_{j}W_{ij}g^{E/I}_j(t), 
\end{equation}
with a firing threshold of $1$ and a reset potential of $0$.
The constant input terms $\mu_i$ were chosen uniformly in the interval $[1.1, 1.2]$ for excitatory neurons, and in the interval $[1, 1.05]$ for inhibitory neurons. 
The membrane time constants for excitatory and inhibitory neurons were set to $\tau_m = 15$~ms and $\tau_m = 10$~ms, respectively. 
The refractory period was fixed at $5$~ms for both excitatory and inhibitory neurons.
Note that although the constant input term is supra-threshold, balanced inputs guaranteed that the average membrane potential is sub-threshold~\cite{Litwin-Kumar2012}.

The network dynamics is captured by the sum in~\eqref{eq:LIF_model}, which describes the input to neuron $i$ from all other neurons in the network.
The topology of the network is encoded by the weight matrix $W$, i.e., $W_{ij}$ denotes the weight of the connection from neuron $j$ to neuron $i$, where $W_{ij}$ is zero if there is no connection.
Synaptic inputs are modeled by $g^{E/I}_j(t)$, which is increased step-wise instantaneously after a presynaptic spike of neuron $j$ ($g_j^{E/I}\rightarrow g_j^{E/I} +1$) and then decays exponentially according to:
\begin{equation}\label{eq:synapse}
    \tau_{E/I} \dfrac{d g_j^{E/I}}{dt} = - g_j^{E/I}(t),
\end{equation}
with time constants $\tau_E =3$ ms for an excitatory interaction, and $\tau_I =2$ ms if the presynaptic neuron is inhibitory.

Equation~\eqref{eq:LIF_model} can be rewritten in matrix notation as:
\begin{equation} \label{eq:LIF_modelMat}
 \dfrac{d\mathbf{V}(t)}{dt} = T^{-1}[\boldsymbol{\mu} - \mathbf{V}] + W \mathbf{g}(t),
\end{equation}
where $T = \text{diag}(\tau_i)$ and 
$\mathbf{V} = [V_1, \ldots, V_N]^T$ and the vectors
$\mathbf{V}, \boldsymbol{\mu}$, and $\mathbf{g}$ are $N \times 1$ vectors.
The spectral analyses in the main text (eigenvalues and Schur vectors) refer to the weight matrix $W$. Different network topologies correspond to different weight matrices $W$, as explained below.

In all simulations, the ratio of excitatory to inhibitory neurons was fixed to be $4:1$. 
Unless otherwise stated, the networks comprise $N =2000$ units ($1600$ excitatory, $400$ inhibitory) and were simulated over $20$ seconds to calculate the statistics reported.
The time step for all simulations was $0.1$ ms.
All simulations were run in MATLAB (version 2011b or later).

\subsection{Network Topologies and Weight matrices}
Different network topologies were simulated using different $W$ matrices but maintaining a general balanced network. If not indicated differently, all parameters below correspond to the case $N=2000$.

\subsubsection{Networks with unclustered balanced connections}
Unless otherwise stated, unclustered connections between neurons were drawn uniformly at random according to the probabilities $p^{EE} =0.2$ and $p^{EI} = p^{IE} = p^{II} = 0.5$, where the first superscript denotes the destination and the second superscript denotes the origin of the synaptic connection, and $E$, $I$ stand for an excitatory or inhibitory neuron, respectively.
Synaptic weights between inhibitory neurons had always the weight $w^{II} = -0.0297$.
The other weight parameters were set to $w^{EI} = -0.0297$, $w^{IE} = 0.0074$ and $w^{EE} = 0.0156$, except where indicated differently in the scenarios described below.

\subsubsection{Networks with clustered excitatory-to-excitatory connections}
First, LIF networks with clustered excitatory neurons were constructed according to the protocol in Ref.~\cite{Litwin-Kumar2012}.
The connection probabilities between excitatory neurons were changed so that neurons had more connections within its group than to neurons outside the group.
The average number of connections was kept the same as for an unclustered balanced network ($p^{EE} = 0.2$) while varying the ratio $R_{EE} = p_{in}/p_{out}$, where $p_{in}$ and $p_{out}$ are the probabilities of connectivity within a group and between groups, respectively. 
$R_{EE} = 1$ corresponds to the unclustered balanced network. 

Second, we generated LIF networks where the connectivity between all excitatory neurons was uniform ($p^{EE}=0.2, \, R_{EE} = 1$) but the weights were varied by changing the ratio $W_{EE} = w_{in}^{EE} / w_{out}^{EE}$, where $w_{in}^{EE}$ refers to synaptic weights within the same group and $w_{out}^{EE}$ refers to synaptic weights to neurons in other groups. 
The average synaptic weight between excitatory neurons was kept at $w^{EE} = 0.0156$, as for the unclustered case.
Strictly speaking, this scheme does not correspond to a clustering in a topological sense, but rather to a clustering of synaptic weights between excitatory neurons.
However, in terms of the dynamics, both E-E clustered network variants 
(adjusting probabilities or weights) lead to effectively the same results, as discussed in the text.
Unless otherwise stated, the excitatory neurons were divided into 20 groups of 80 neurons each.

As the weights in balanced networks should scale with $1/\sqrt{N}$~\cite{Vreeswijk1998, Litwin-Kumar2012}, when varying the network size (see Fig.~\ref{fig:2}A)  the parameter settings were kept the same but all weights were scaled accordingly, e.g., for $N=1000$ the weights were multiplied by $\sqrt{2}$ compared to those of the network with $N=2000$.

\subsubsection{Networks with excitatory-to-inhibitory feedback loop}
To construct LIF networks with co-clustered excitatory and inhibitory neurons, we kept uniform excitatory-to-excitatory and inhibitory-to-inhibitory couplings while varying the ratios $W_{IE} = w^{IE}_{in} / w^{IE}_{out}$, $W_{EI} = (w^{EI}_{in} / w^{EI}_{out})^{-1}$, or the analogous ratios of connections probabilities $R_{IE}, R_{EI}$.
The subscript `in' indicates connections within the excitatory/inhibitory pair.
Note that $W_{EI}, R_{EI}$ are defined by the inverse in/out ratio, since they describe an inhibitory effect.
To modulate the co-clustered excitatory-to-inhibitory network dynamics, we vary the ratios 
$W_{IE}, W_{EI}$, while keeping the average weights constant. 
For instance, in the example shown in Figure \ref{fig:4_part2}, we kept $R_{EI} = R_{IE} =2$ while simultaneously varying the ratios $W_{IE}= W_{EI}$ between 1 and 5. 
For all networks, we divided the network into 20 groups with 80 excitatory and 20 inhibitory neurons each.

\subsubsection{Networks with small-world connectivity}
Networks with small-world connectivity between excitatory neurons were constructed as follows.
Excitatory neurons ($N=1600$) were connected within a 40-nearest neighbor ring with probability $p^{EE}_{in}$, i.e., neurons $i$  was connected to neuron $j$ with probability $p^{EE}_{in}$ if $|i-j|\leq40$ subject to periodic boundary conditions.
The connection probability outside this neighborhood was set to $p^{EE}_{out}$.
As in previous models, the average connectivity was kept constant at $p^{EE}=0.2$ 
while varying the ratio $R_{EE}^{SW} = p^{EE}_{in}/p^{EE}_{out}$.
Hence, by increasing $R_{EE}^{SW}$ we increase the amount of `backbone' connectivity through the (stochastic) 40 nearest-neighbor cycle, while decreasing the number of connections to neurons elsewhere in the network.
This construction may thus also be interpreted as an overlapping clustering and is effectively a variant of the small-world scheme introduced by Watts and Strogatz \cite{Watts1998}.

\subsubsection{Networks with scale-free connectivity}
LIF networks with a scale-free topology for the excitatory connections were created by a preferential attachment process~\cite{Barabasi1999} using algorithm $5$ in \cite{Batagelj2005} with parameter $d=64$.
The weight matrix was created iteratively by adding one neuron at a time, such that the probability that it is connected to an existing node is proportional to the  degree of that node (at the current state), \textit{i.e.},
the earlier a node is added, the larger its final degree will be. The resulting degree of the network tends to a power-law distribution \cite{Barabasi1999}.
The value $d = 64$ was chosen to avoid saturation of the dynamics of the network, and this leads to an excitatory-excitatory matrix with $35\%$ of the edges of our other examples.

\subsubsection{Network with hierarchical excitatory connections}
We simulated LIF networks where excitatory neurons belonged to a hierarchy of groups (Figure \ref{fig:5_hier}) by dividing the excitatory neurons into nested clusters.
We implemented two layers in the hierarchy: the excitatory neurons were divided into $16$ groups, and each of these groups was then subdivided into $2$ more groups to give a total of $32$ subgroups.
This was achieved by varying the ratios $R_{EE}^{top} = p^{EE}_{\text{group}}/p^{EE}_{out}$, $R_{EE}^{sub} = p^{EE}_{\text{subgroup}}/p^{EE}_{\text{group}}$ while keeping the  average connectivity between excitatory neurons constant.
The example in Figure \ref{fig:5_hier} corresponds to $R_{EE}^{top} = 1.45$ and $R_{EE}^{sub} =3.7$. The weight within the subgroups was set at $w^{EE}_{sub} = 0.0163$.
All other connections remained unchanged.

\subsection{Quantifying SSA dynamics from spike-train LIF simulations}
To evaluate the extent to which a network displays slow-switching assembly dynamics, we used the following two spike-rate variability measures.

Given a partition of the neurons into $c$ groups, we compute the average spiking frequency $f_i(t)$ of the neurons in each cluster $i \in \{1,\ldots,c\}$ over non-overlapping windows of $100$ms.
For each time window, we obtain the firing rate vector $\mathbf{f}(t) = [f_1(t), \ldots, f_c(t)]^T$ of which we compute the standard deviation $\sigma(t)$.
The standard deviations are then averaged over the duration of the simulation:
\begin{equation}
S =\frac{1}{T}\sum_{t=1}^T \sigma(t),
\end{equation}
where $T$ is the total number of time windows in the simulation.
We then obtain a boot-strapped expectation $ \langle S^\text{shuff} \rangle$ computed by reshuffling neurons at random into groups of the same sizes as those in the partition.
The spike rate variability score is then defined as:
\begin{align}
\label{eq:S_hat}
  \widehat S = S - \langle S^\text{shuff} \rangle,
\end{align}
where the average $\langle S^\text{shuff} \rangle$ is obtained over 10 random reshufflings of the neurons. If the network fires uniformly (with no localized patterns), $\widehat S$ is low; whereas  $\widehat S$ increases if the network displays heterogeneous activity aligned with the partition under investigation.

As explained in the text, we introduced a second spike-train variability measure to quantify the variations of the group firing-rates across time. This complementary measure allows us to discern scenarios in which there is a group of neurons dominating the firing (thus leading to a large variation across groups), but no switching between groups. To quantify these effects we use an analogous measure to $\widehat S$ above.

Given a partition of the neurons into $c$ groups, we compute the average spiking frequency $f_i(t)$ of the neurons in each cluster $i \in \{1,\ldots,c\}$ over non-overlapping windows of $100$ms.
For each group $i$,  we compute the standard deviation $\sigma_T(i)$ of the vector of coarse-grained firing rates across time $\mathbf{f_i} = [f_i(t_1), \ldots, f_i(t_N)]$, where $t_k$ stands for the $k$th time bin and $i=1,\ldots, c$.  We then average over groups to obtain:
\begin{align}
S_T =\frac{1}{c}\sum_{i=1}^c \sigma_T(i).
\end{align}
As above, we obtain a boot-strapped expectation $ \langle S_T^\text{shuff} \rangle$ by reshuffling neurons at random into groups of the same sizes as those in the partition.
The spike rate variability score over time is then defined as:
\begin{align}
\label{eq:S_That}
  \widehat S_T = S_T - \langle S_T^\text{shuff} \rangle,
\end{align}
where the average $\langle S_T^\text{shuff} \rangle$ is obtained over 10 random reshufflings of the neurons. 
Again, if the network fires homogeneously (with no localized patterns in time), $\widehat S_T$ is low; whereas  
$\widehat S_T$ increases if the network displays heterogeneous firing rates over time. 

In Figure S1, we show the behavior of $\widehat S_T$ for the examples discussed in the main manuscript. As expected, for all cases where SSA dynamics is present, both measures $\widehat S$ and $\widehat S_T$ behave consistently. On the other hand,  $\widehat S_T$ detects the end of the SSA region when, through increased clustering, the dynamics gets localized on one cell assembly, as shown in Figure~\ref{fig:S2}.

\subsection{Measuring alignment of LIF dynamics with the Schur vectors of the weight matrix: the principal angle}

First, we find the dominant firing patterns in the network dynamics. We perform simulations of the LIF network and obtain the firing rates of every neuron in $250$ ms bins to generate an $N \times T$ matrix, where $N$ is the number of neurons and $T$ is the number of bins. On this matrix, we perform a \textit{principal component analysis} (PCA) and select the first $c-1$ principal components $\{\mathbf{p_i}\}_{i=1}^{c-1}$. This set of $N$-dimensional vectors captures most of the variability observed in the simulated dynamics.

We then assess how aligned the $c-1$ principal components are with the dominant Schur vectors of the weight matrix of the network.
More precisely, we compute the Schur decomposition of the weight matrix $W$; keep the $c-1$ dominant Schur vectors $\{\mathbf{u_i}\}_{i=1}^{c-1}$ associated with the eigenvalues with largest real part; and then compute the (first) principal or canonical angle $\theta$ between the subspaces spanned by the two sets of vectors $\mathcal P = \text{span}\{\mathbf{p_i}\}$ and $\mathcal U = \text{span}\{\mathbf{u_i}\}$~\cite{Golub1996,Stewart2001}:
\begin{equation}
\label{eq:theta}
  \cos(\theta) = \max \left\lbrace \left. \dfrac{\mathbf{u}^T\mathbf{p}}{\|\mathbf{u}\| \| \mathbf{p} \|}  \right| \mathbf{u}\in \mathcal U \:\:\mathbf{p} \in \mathcal P\right\rbrace,
\end{equation}
The first principal angle measures how `close' the observed firing patterns are to the dominant modes computed solely from the weight matrix. If $\cos(\theta) \approx 1$, there is a large alignment between the span of both subspaces.

\bibliography{./literature}

\begin{thebibliography}{74}%
\makeatletter
\providecommand \@ifxundefined [1]{%
 \@ifx{#1\undefined}
}%
\providecommand \@ifnum [1]{%
 \ifnum #1\expandafter \@firstoftwo
 \else \expandafter \@secondoftwo
 \fi
}%
\providecommand \@ifx [1]{%
 \ifx #1\expandafter \@firstoftwo
 \else \expandafter \@secondoftwo
 \fi
}%
\providecommand \natexlab [1]{#1}%
\providecommand \enquote  [1]{``#1''}%
\providecommand \bibnamefont  [1]{#1}%
\providecommand \bibfnamefont [1]{#1}%
\providecommand \citenamefont [1]{#1}%
\providecommand \href@noop [0]{\@secondoftwo}%
\providecommand \href [0]{\begingroup \@sanitize@url \@href}%
\providecommand \@href[1]{\@@startlink{#1}\@@href}%
\providecommand \@@href[1]{\endgroup#1\@@endlink}%
\providecommand \@sanitize@url [0]{\catcode `\\12\catcode `\$12\catcode
  `\&12\catcode `\#12\catcode `\^12\catcode `\_12\catcode `\%12\relax}%
\providecommand \@@startlink[1]{}%
\providecommand \@@endlink[0]{}%
\providecommand \url  [0]{\begingroup\@sanitize@url \@url }%
\providecommand \@url [1]{\endgroup\@href {#1}{\urlprefix }}%
\providecommand \urlprefix  [0]{URL }%
\providecommand \Eprint [0]{\href }%
\providecommand \doibase [0]{http://dx.doi.org/}%
\providecommand \selectlanguage [0]{\@gobble}%
\providecommand \bibinfo  [0]{\@secondoftwo}%
\providecommand \bibfield  [0]{\@secondoftwo}%
\providecommand \translation [1]{[#1]}%
\providecommand \BibitemOpen [0]{}%
\providecommand \bibitemStop [0]{}%
\providecommand \bibitemNoStop [0]{.\EOS\space}%
\providecommand \EOS [0]{\spacefactor3000\relax}%
\providecommand \BibitemShut  [1]{\csname bibitem#1\endcsname}%
\let\auto@bib@innerbib\@empty
\bibitem [{\citenamefont {Buzsaki}(2004)}]{Buzsaki2004}%
  \BibitemOpen
  \bibfield  {author} {\bibinfo {author} {\bibfnamefont {G.}~\bibnamefont
  {Buzsaki}},\ }\href {\doibase 10.1038/nn1233} {\bibfield  {journal} {\bibinfo
   {journal} {Nature Neuroscience}\ }\textbf {\bibinfo {volume} {7}},\ \bibinfo
  {pages} {446} (\bibinfo {year} {2004})}\BibitemShut {NoStop}%
\bibitem [{\citenamefont {Du}\ \emph {et~al.}(2011)\citenamefont {Du},
  \citenamefont {Blanche}, \citenamefont {Harrison}, \citenamefont {Lester},\
  and\ \citenamefont {Masmanidis}}]{Du2011}%
  \BibitemOpen
  \bibfield  {author} {\bibinfo {author} {\bibfnamefont {J.}~\bibnamefont
  {Du}}, \bibinfo {author} {\bibfnamefont {T.~J.}\ \bibnamefont {Blanche}},
  \bibinfo {author} {\bibfnamefont {R.~R.}\ \bibnamefont {Harrison}}, \bibinfo
  {author} {\bibfnamefont {H.~A.}\ \bibnamefont {Lester}}, \ and\ \bibinfo
  {author} {\bibfnamefont {S.~C.}\ \bibnamefont {Masmanidis}},\ }\href
  {\doibase 10.1371/journal.pone.0026204} {\bibfield  {journal} {\bibinfo
  {journal} {PLoS ONE}\ }\textbf {\bibinfo {volume} {6}} (\bibinfo {year}
  {2011}),\ 10.1371/journal.pone.0026204}\BibitemShut {NoStop}%
\bibitem [{\citenamefont {Ahrens}\ \emph {et~al.}(2013)\citenamefont {Ahrens},
  \citenamefont {Orger}, \citenamefont {Robson}, \citenamefont {Li},\ and\
  \citenamefont {Keller}}]{Ahrens2013}%
  \BibitemOpen
  \bibfield  {author} {\bibinfo {author} {\bibfnamefont {M.~B.}\ \bibnamefont
  {Ahrens}}, \bibinfo {author} {\bibfnamefont {M.~B.}\ \bibnamefont {Orger}},
  \bibinfo {author} {\bibfnamefont {D.~N.}\ \bibnamefont {Robson}}, \bibinfo
  {author} {\bibfnamefont {J.~M.}\ \bibnamefont {Li}}, \ and\ \bibinfo {author}
  {\bibfnamefont {P.~J.}\ \bibnamefont {Keller}},\ }\href {\doibase
  10.1038/NMETH.2434} {\bibfield  {journal} {\bibinfo  {journal} {Nature
  Methods}\ }\textbf {\bibinfo {volume} {10}},\ \bibinfo {pages} {413+}
  (\bibinfo {year} {2013})}\BibitemShut {NoStop}%
\bibitem [{\citenamefont {Buzsaki}(2010)}]{Buzsaki2010}%
  \BibitemOpen
  \bibfield  {author} {\bibinfo {author} {\bibfnamefont {G.}~\bibnamefont
  {Buzsaki}},\ }\href {\doibase 10.1016/j.neuron.2010.09.023} {\bibfield
  {journal} {\bibinfo  {journal} {Neuron}\ }\textbf {\bibinfo {volume} {68}},\
  \bibinfo {pages} {362} (\bibinfo {year} {2010})}\BibitemShut {NoStop}%
\bibitem [{\citenamefont {Hebb}(1949)}]{Hebb1949}%
  \BibitemOpen
  \bibfield  {author} {\bibinfo {author} {\bibfnamefont {D.~O.}\ \bibnamefont
  {Hebb}},\ }\href@noop {} {\emph {\bibinfo {title} {{T}he {O}rganization of
  {B}ehavior}}}\ (\bibinfo  {publisher} {JohnWiley \& Sons},\ \bibinfo {year}
  {1949})\BibitemShut {NoStop}%
\bibitem [{\citenamefont {Harris}(2005)}]{Harris2005}%
  \BibitemOpen
  \bibfield  {author} {\bibinfo {author} {\bibfnamefont {K.}~\bibnamefont
  {Harris}},\ }\href {\doibase 10.1038/nrn1669} {\bibfield  {journal} {\bibinfo
   {journal} {Nature Reviews Neuroscience}\ }\textbf {\bibinfo {volume} {6}},\
  \bibinfo {pages} {399} (\bibinfo {year} {2005})}\BibitemShut {NoStop}%
\bibitem [{\citenamefont {Song}\ \emph {et~al.}(2005)\citenamefont {Song},
  \citenamefont {Sjostrom}, \citenamefont {Reigl}, \citenamefont {Nelson},\
  and\ \citenamefont {Chklovskii}}]{Song2005}%
  \BibitemOpen
  \bibfield  {author} {\bibinfo {author} {\bibfnamefont {S.}~\bibnamefont
  {Song}}, \bibinfo {author} {\bibfnamefont {P.}~\bibnamefont {Sjostrom}},
  \bibinfo {author} {\bibfnamefont {M.}~\bibnamefont {Reigl}}, \bibinfo
  {author} {\bibfnamefont {S.}~\bibnamefont {Nelson}}, \ and\ \bibinfo {author}
  {\bibfnamefont {D.}~\bibnamefont {Chklovskii}},\ }\href {\doibase
  10.1371/journal.pbio.0030068} {\bibfield  {journal} {\bibinfo  {journal}
  {Plos Biology}\ }\textbf {\bibinfo {volume} {3}},\ \bibinfo {pages} {507}
  (\bibinfo {year} {2005})}\BibitemShut {NoStop}%
\bibitem [{\citenamefont {Perin}\ \emph {et~al.}(2011)\citenamefont {Perin},
  \citenamefont {Berger},\ and\ \citenamefont {Markram}}]{Perin2011}%
  \BibitemOpen
  \bibfield  {author} {\bibinfo {author} {\bibfnamefont {R.}~\bibnamefont
  {Perin}}, \bibinfo {author} {\bibfnamefont {T.~K.}\ \bibnamefont {Berger}}, \
  and\ \bibinfo {author} {\bibfnamefont {H.}~\bibnamefont {Markram}},\ }\href
  {\doibase 10.1073/pnas.1016051108} {\bibfield  {journal} {\bibinfo  {journal}
  {Proceedings of the National Academy of Sciences of the United States of
  America}\ }\textbf {\bibinfo {volume} {108}},\ \bibinfo {pages} {5419}
  (\bibinfo {year} {2011})}\BibitemShut {NoStop}%
\bibitem [{\citenamefont {Yoshimura}\ \emph {et~al.}(2005)\citenamefont
  {Yoshimura}, \citenamefont {Dantzker},\ and\ \citenamefont
  {Callaway}}]{Yoshimura2005}%
  \BibitemOpen
  \bibfield  {author} {\bibinfo {author} {\bibfnamefont {Y.}~\bibnamefont
  {Yoshimura}}, \bibinfo {author} {\bibfnamefont {J.}~\bibnamefont {Dantzker}},
  \ and\ \bibinfo {author} {\bibfnamefont {E.}~\bibnamefont {Callaway}},\
  }\href {\doibase 10.1038/nature03291} {\bibfield  {journal} {\bibinfo
  {journal} {Nature}\ }\textbf {\bibinfo {volume} {433}},\ \bibinfo {pages}
  {868} (\bibinfo {year} {2005})}\BibitemShut {NoStop}%
\bibitem [{\citenamefont {Otsuka}\ and\ \citenamefont
  {Kawaguchi}(2011)}]{Otsuka2011}%
  \BibitemOpen
  \bibfield  {author} {\bibinfo {author} {\bibfnamefont {T.}~\bibnamefont
  {Otsuka}}\ and\ \bibinfo {author} {\bibfnamefont {Y.}~\bibnamefont
  {Kawaguchi}},\ }\href {\doibase 10.1523/JNEUROSCI.5795-10.2011} {\bibfield
  {journal} {\bibinfo  {journal} {Journal of Neuroscience}\ }\textbf {\bibinfo
  {volume} {31}},\ \bibinfo {pages} {3862} (\bibinfo {year}
  {2011})}\BibitemShut {NoStop}%
\bibitem [{\citenamefont {Ko}\ \emph {et~al.}(2011)\citenamefont {Ko},
  \citenamefont {Hofer}, \citenamefont {Pichler}, \citenamefont {Buchanan},
  \citenamefont {Sjoestroem},\ and\ \citenamefont {Mrsic-Flogel}}]{Ko2011}%
  \BibitemOpen
  \bibfield  {author} {\bibinfo {author} {\bibfnamefont {H.}~\bibnamefont
  {Ko}}, \bibinfo {author} {\bibfnamefont {S.~B.}\ \bibnamefont {Hofer}},
  \bibinfo {author} {\bibfnamefont {B.}~\bibnamefont {Pichler}}, \bibinfo
  {author} {\bibfnamefont {K.~A.}\ \bibnamefont {Buchanan}}, \bibinfo {author}
  {\bibfnamefont {P.~J.}\ \bibnamefont {Sjoestroem}}, \ and\ \bibinfo {author}
  {\bibfnamefont {T.~D.}\ \bibnamefont {Mrsic-Flogel}},\ }\href {\doibase
  10.1038/nature09880} {\bibfield  {journal} {\bibinfo  {journal} {Nature}\
  }\textbf {\bibinfo {volume} {473}},\ \bibinfo {pages} {87} (\bibinfo {year}
  {2011})}\BibitemShut {NoStop}%
\bibitem [{\citenamefont {Harris}\ and\ \citenamefont
  {Mrsic-Flogel}(2013)}]{Harris2013}%
  \BibitemOpen
  \bibfield  {author} {\bibinfo {author} {\bibfnamefont {K.~D.}\ \bibnamefont
  {Harris}}\ and\ \bibinfo {author} {\bibfnamefont {T.~D.}\ \bibnamefont
  {Mrsic-Flogel}},\ }\href {\doibase 10.1038/nature12654} {\bibfield  {journal}
  {\bibinfo  {journal} {Nature}\ }\textbf {\bibinfo {volume} {503}},\ \bibinfo
  {pages} {51} (\bibinfo {year} {2013})}\BibitemShut {NoStop}%
\bibitem [{\citenamefont {Lefort}\ \emph {et~al.}(2009)\citenamefont {Lefort},
  \citenamefont {Tomm}, \citenamefont {Sarria},\ and\ \citenamefont
  {Petersen}}]{Lefort2009}%
  \BibitemOpen
  \bibfield  {author} {\bibinfo {author} {\bibfnamefont {S.}~\bibnamefont
  {Lefort}}, \bibinfo {author} {\bibfnamefont {C.}~\bibnamefont {Tomm}},
  \bibinfo {author} {\bibfnamefont {J.~C.~F.}\ \bibnamefont {Sarria}}, \ and\
  \bibinfo {author} {\bibfnamefont {C.~C.~H.}\ \bibnamefont {Petersen}},\
  }\href {\doibase 10.1016/j.neuron.2008.12.020} {\bibfield  {journal}
  {\bibinfo  {journal} {Neuron}\ }\textbf {\bibinfo {volume} {61}},\ \bibinfo
  {pages} {301} (\bibinfo {year} {2009})}\BibitemShut {NoStop}%
\bibitem [{\citenamefont {Yassin}\ \emph {et~al.}(2010)\citenamefont {Yassin},
  \citenamefont {Benedetti}, \citenamefont {Jouhanneau}, \citenamefont {Wen},
  \citenamefont {Poulet},\ and\ \citenamefont {Barth}}]{Yassin2010}%
  \BibitemOpen
  \bibfield  {author} {\bibinfo {author} {\bibfnamefont {L.}~\bibnamefont
  {Yassin}}, \bibinfo {author} {\bibfnamefont {B.~L.}\ \bibnamefont
  {Benedetti}}, \bibinfo {author} {\bibfnamefont {J.-S.}\ \bibnamefont
  {Jouhanneau}}, \bibinfo {author} {\bibfnamefont {J.~A.}\ \bibnamefont {Wen}},
  \bibinfo {author} {\bibfnamefont {J.~F.~A.}\ \bibnamefont {Poulet}}, \ and\
  \bibinfo {author} {\bibfnamefont {A.~L.}\ \bibnamefont {Barth}},\ }\href
  {\doibase 10.1016/j.neuron.2010.11.029} {\bibfield  {journal} {\bibinfo
  {journal} {Neuron}\ }\textbf {\bibinfo {volume} {68}},\ \bibinfo {pages}
  {1043} (\bibinfo {year} {2010})}\BibitemShut {NoStop}%
\bibitem [{\citenamefont {Shimono}\ and\ \citenamefont
  {Beggs}(2014)}]{Shimono2014}%
  \BibitemOpen
  \bibfield  {author} {\bibinfo {author} {\bibfnamefont {M.}~\bibnamefont
  {Shimono}}\ and\ \bibinfo {author} {\bibfnamefont {J.~M.}\ \bibnamefont
  {Beggs}},\ }\href {\doibase 10.1093/cercor/bhu252} {\bibfield  {journal}
  {\bibinfo  {journal} {Cerebral Cortex}\ } (\bibinfo {year} {2014}),\
  10.1093/cercor/bhu252}\BibitemShut {NoStop}%
\bibitem [{\citenamefont {McGinley}\ and\ \citenamefont
  {Westbrook}(2013)}]{McGinley2013}%
  \BibitemOpen
  \bibfield  {author} {\bibinfo {author} {\bibfnamefont {M.~J.}\ \bibnamefont
  {McGinley}}\ and\ \bibinfo {author} {\bibfnamefont {G.~L.}\ \bibnamefont
  {Westbrook}},\ }\href {\doibase 10.1073/pnas.1303813110} {\bibfield
  {journal} {\bibinfo  {journal} {Proceedings of the National Academy of
  Sciences of the United States of America}\ }\textbf {\bibinfo {volume}
  {110}},\ \bibinfo {pages} {16193} (\bibinfo {year} {2013})}\BibitemShut
  {NoStop}%
\bibitem [{\citenamefont {Savic}\ \emph {et~al.}(2000)\citenamefont {Savic},
  \citenamefont {Gulyas}, \citenamefont {Larsson},\ and\ \citenamefont
  {Roland}}]{Savic2000}%
  \BibitemOpen
  \bibfield  {author} {\bibinfo {author} {\bibfnamefont {I.}~\bibnamefont
  {Savic}}, \bibinfo {author} {\bibfnamefont {B.}~\bibnamefont {Gulyas}},
  \bibinfo {author} {\bibfnamefont {M.}~\bibnamefont {Larsson}}, \ and\
  \bibinfo {author} {\bibfnamefont {P.}~\bibnamefont {Roland}},\ }\href
  {\doibase 10.1016/S0896-6273(00)81209-X} {\bibfield  {journal} {\bibinfo
  {journal} {Neuron}\ }\textbf {\bibinfo {volume} {26}},\ \bibinfo {pages}
  {735} (\bibinfo {year} {2000})}\BibitemShut {NoStop}%
\bibitem [{\citenamefont {Felleman}\ and\ \citenamefont
  {Van~Essen}(1991)}]{Felleman1991}%
  \BibitemOpen
  \bibfield  {author} {\bibinfo {author} {\bibfnamefont {D.~J.}\ \bibnamefont
  {Felleman}}\ and\ \bibinfo {author} {\bibfnamefont {D.~C.}\ \bibnamefont
  {Van~Essen}},\ }\href {\doibase {10.1093/cercor/1.1.1}} {\bibfield  {journal}
  {\bibinfo  {journal} {Cerebral Cortex}\ }\textbf {\bibinfo {volume} {1}},\
  \bibinfo {pages} {1} (\bibinfo {year} {1991})}\BibitemShut {NoStop}%
\bibitem [{\citenamefont {Ito}\ \emph {et~al.}(2013)\citenamefont {Ito},
  \citenamefont {Masuda}, \citenamefont {Shinomiya}, \citenamefont {Endo},\
  and\ \citenamefont {Ito}}]{Ito2013}%
  \BibitemOpen
  \bibfield  {author} {\bibinfo {author} {\bibfnamefont {M.}~\bibnamefont
  {Ito}}, \bibinfo {author} {\bibfnamefont {N.}~\bibnamefont {Masuda}},
  \bibinfo {author} {\bibfnamefont {K.}~\bibnamefont {Shinomiya}}, \bibinfo
  {author} {\bibfnamefont {K.}~\bibnamefont {Endo}}, \ and\ \bibinfo {author}
  {\bibfnamefont {K.}~\bibnamefont {Ito}},\ }\href {\doibase
  10.1016/j.cub.2013.03.015} {\bibfield  {journal} {\bibinfo  {journal}
  {Current Biology}\ }\textbf {\bibinfo {volume} {23}},\ \bibinfo {pages} {644}
  (\bibinfo {year} {2013})}\BibitemShut {NoStop}%
\bibitem [{\citenamefont {J~Pavlides}\ \emph {et~al.}(1988)\citenamefont
  {J~Pavlides}, \citenamefont {Greenstein}, \citenamefont {Grudman},\ and\
  \citenamefont {J}}]{Pavlides1988}%
  \BibitemOpen
  \bibfield  {author} {\bibinfo {author} {\bibfnamefont {C.}~\bibnamefont
  {J~Pavlides}}, \bibinfo {author} {\bibfnamefont {Y.}~\bibnamefont
  {Greenstein}}, \bibinfo {author} {\bibfnamefont {M.}~\bibnamefont {Grudman}},
  \ and\ \bibinfo {author} {\bibfnamefont {W.}~\bibnamefont {J}},\ }\href
  {\doibase 10.1016/0006-8993(88)91499-0} {\bibfield  {journal} {\bibinfo
  {journal} {Brain Research}\ }\textbf {\bibinfo {volume} {439}},\ \bibinfo
  {pages} {383} (\bibinfo {year} {1988})}\BibitemShut {NoStop}%
\bibitem [{\citenamefont {Hyman}\ \emph {et~al.}(2003)\citenamefont {Hyman},
  \citenamefont {Wyble}, \citenamefont {Goyal}, \citenamefont {Rossi},\ and\
  \citenamefont {Hasselmo}}]{Hyman2003}%
  \BibitemOpen
  \bibfield  {author} {\bibinfo {author} {\bibfnamefont {J.}~\bibnamefont
  {Hyman}}, \bibinfo {author} {\bibfnamefont {B.}~\bibnamefont {Wyble}},
  \bibinfo {author} {\bibfnamefont {V.}~\bibnamefont {Goyal}}, \bibinfo
  {author} {\bibfnamefont {C.}~\bibnamefont {Rossi}}, \ and\ \bibinfo {author}
  {\bibfnamefont {M.}~\bibnamefont {Hasselmo}},\ }\href@noop {} {\bibfield
  {journal} {\bibinfo  {journal} {Journal of Neuroscience}\ }\textbf {\bibinfo
  {volume} {23}},\ \bibinfo {pages} {11725} (\bibinfo {year}
  {2003})}\BibitemShut {NoStop}%
\bibitem [{\citenamefont {Litwin-Kumar}\ and\ \citenamefont
  {Doiron}(2012)}]{Litwin-Kumar2012}%
  \BibitemOpen
  \bibfield  {author} {\bibinfo {author} {\bibfnamefont {A.}~\bibnamefont
  {Litwin-Kumar}}\ and\ \bibinfo {author} {\bibfnamefont {B.}~\bibnamefont
  {Doiron}},\ }\href {http://dx.doi.org/10.1038/nn.3220} {\bibfield  {journal}
  {\bibinfo  {journal} {Nature Neuroscince}\ }\textbf {\bibinfo {volume}
  {15}},\ \bibinfo {pages} {1498} (\bibinfo {year} {2012})}\BibitemShut
  {NoStop}%
\bibitem [{\citenamefont {van Vreeswijk}\ and\ \citenamefont
  {Sompolinsky}(1998)}]{Vreeswijk1998}%
  \BibitemOpen
  \bibfield  {author} {\bibinfo {author} {\bibfnamefont {C.}~\bibnamefont {van
  Vreeswijk}}\ and\ \bibinfo {author} {\bibfnamefont {H.}~\bibnamefont
  {Sompolinsky}},\ }\href@noop {} {\bibfield  {journal} {\bibinfo  {journal}
  {Neural Computation}\ }\textbf {\bibinfo {volume} {10}},\ \bibinfo {pages}
  {1321} (\bibinfo {year} {1998})}\BibitemShut {NoStop}%
\bibitem [{\citenamefont {Murphy}\ and\ \citenamefont
  {Miller}(2009)}]{Murphy2009}%
  \BibitemOpen
  \bibfield  {author} {\bibinfo {author} {\bibfnamefont {B.~K.}\ \bibnamefont
  {Murphy}}\ and\ \bibinfo {author} {\bibfnamefont {K.~D.}\ \bibnamefont
  {Miller}},\ }\href {\doibase 10.1016/j.neuron.2009.02.005} {\bibfield
  {journal} {\bibinfo  {journal} {Neuron}\ }\textbf {\bibinfo {volume} {61}},\
  \bibinfo {pages} {635 } (\bibinfo {year} {2009})}\BibitemShut {NoStop}%
\bibitem [{\citenamefont {von Luxburg}(2007)}]{Luxburg2007}%
  \BibitemOpen
  \bibfield  {author} {\bibinfo {author} {\bibfnamefont {U.}~\bibnamefont {von
  Luxburg}},\ }\href {\doibase 10.1007/s11222-007-9033-z} {\bibfield  {journal}
  {\bibinfo  {journal} {Statistics and Computing}\ }\textbf {\bibinfo {volume}
  {17}},\ \bibinfo {pages} {395} (\bibinfo {year} {2007})}\BibitemShut
  {NoStop}%
\bibitem [{\citenamefont {Zhang}\ \emph {et~al.}(2013)\citenamefont {Zhang},
  \citenamefont {Rao~Nadakuditi},\ and\ \citenamefont {Newman}}]{Zhang2013}%
  \BibitemOpen
  \bibfield  {author} {\bibinfo {author} {\bibfnamefont {X.}~\bibnamefont
  {Zhang}}, \bibinfo {author} {\bibfnamefont {R.}~\bibnamefont
  {Rao~Nadakuditi}}, \ and\ \bibinfo {author} {\bibfnamefont {M.~E.~J.}\
  \bibnamefont {Newman}},\ }\href {http://arxiv.org/abs/1310.0046v1} {\enquote
  {\bibinfo {title} {{S}pectra of random graphs with community structure and
  arbitrary degrees},}\ } (\bibinfo {year} {2013}),\ \bibinfo {note}
  {arXiv:1310.0046}\BibitemShut {NoStop}%
\bibitem [{\citenamefont {Nadakuditi}\ and\ \citenamefont
  {Newman}(2013)}]{Nadakuditi2013}%
  \BibitemOpen
  \bibfield  {author} {\bibinfo {author} {\bibfnamefont {R.~R.}\ \bibnamefont
  {Nadakuditi}}\ and\ \bibinfo {author} {\bibfnamefont {M.~E.~J.}\ \bibnamefont
  {Newman}},\ }\href {\doibase 10.1103/PhysRevE.87.012803} {\bibfield
  {journal} {\bibinfo  {journal} {Physical Review E}\ }\textbf {\bibinfo
  {volume} {87}},\ \bibinfo {pages} {012803} (\bibinfo {year}
  {2013})}\BibitemShut {NoStop}%
\bibitem [{\citenamefont {Simon}\ and\ \citenamefont {Ando}(1961)}]{Simon1961}%
  \BibitemOpen
  \bibfield  {author} {\bibinfo {author} {\bibfnamefont {H.~A.}\ \bibnamefont
  {Simon}}\ and\ \bibinfo {author} {\bibfnamefont {A.}~\bibnamefont {Ando}},\
  }\href {http://www.jstor.org/stable/1909285} {\bibfield  {journal} {\bibinfo
  {journal} {Econometrica}\ }\textbf {\bibinfo {volume} {29}},\ \bibinfo
  {pages} {111} (\bibinfo {year} {1961})}\BibitemShut {NoStop}%
\bibitem [{\citenamefont {Gal\'{a}n}(2008)}]{Galan2008}%
  \BibitemOpen
  \bibfield  {author} {\bibinfo {author} {\bibfnamefont {R.~F.}\ \bibnamefont
  {Gal\'{a}n}},\ }\href {\doibase 10.1371/journal.pone.0002148} {\bibfield
  {journal} {\bibinfo  {journal} {PLoS ONE}\ }\textbf {\bibinfo {volume} {3}},\
  \bibinfo {pages} {e2148} (\bibinfo {year} {2008})}\BibitemShut {NoStop}%
\bibitem [{\citenamefont {Delvenne}\ \emph {et~al.}(2010)\citenamefont
  {Delvenne}, \citenamefont {Yaliraki},\ and\ \citenamefont
  {Barahona}}]{Delvenne2010}%
  \BibitemOpen
  \bibfield  {author} {\bibinfo {author} {\bibfnamefont {J.-C.}\ \bibnamefont
  {Delvenne}}, \bibinfo {author} {\bibfnamefont {S.~N.}\ \bibnamefont
  {Yaliraki}}, \ and\ \bibinfo {author} {\bibfnamefont {M.}~\bibnamefont
  {Barahona}},\ }\href {\doibase 10.1073/pnas.0903215107} {\bibfield  {journal}
  {\bibinfo  {journal} {Proceedings of the National Academy of Sciences}\
  }\textbf {\bibinfo {volume} {107}},\ \bibinfo {pages} {12755} (\bibinfo
  {year} {2010})}\BibitemShut {NoStop}%
\bibitem [{\citenamefont {Schaub}\ \emph {et~al.}(2012)\citenamefont {Schaub},
  \citenamefont {Delvenne}, \citenamefont {Yaliraki},\ and\ \citenamefont
  {Barahona}}]{Schaub2012}%
  \BibitemOpen
  \bibfield  {author} {\bibinfo {author} {\bibfnamefont {M.~T.}\ \bibnamefont
  {Schaub}}, \bibinfo {author} {\bibfnamefont {J.-C.}\ \bibnamefont
  {Delvenne}}, \bibinfo {author} {\bibfnamefont {S.~N.}\ \bibnamefont
  {Yaliraki}}, \ and\ \bibinfo {author} {\bibfnamefont {M.}~\bibnamefont
  {Barahona}},\ }\href {\doibase 10.1371/journal.pone.0032210} {\bibfield
  {journal} {\bibinfo  {journal} {PLoS ONE}\ }\textbf {\bibinfo {volume} {7}},\
  \bibinfo {pages} {e32210} (\bibinfo {year} {2012})}\BibitemShut {NoStop}%
\bibitem [{\citenamefont {Delvenne}\ \emph {et~al.}(2013)\citenamefont
  {Delvenne}, \citenamefont {Schaub}, \citenamefont {Yaliraki},\ and\
  \citenamefont {Barahona}}]{Delvenne2013}%
  \BibitemOpen
  \bibfield  {author} {\bibinfo {author} {\bibfnamefont {J.-C.}\ \bibnamefont
  {Delvenne}}, \bibinfo {author} {\bibfnamefont {M.~T.}\ \bibnamefont
  {Schaub}}, \bibinfo {author} {\bibfnamefont {S.~N.}\ \bibnamefont
  {Yaliraki}}, \ and\ \bibinfo {author} {\bibfnamefont {M.}~\bibnamefont
  {Barahona}},\ }in\ \href {\doibase 10.1007/978-1-4614-6729-8_11} {\emph
  {\bibinfo {booktitle} {Dynamics On and Of Complex Networks, Volume 2}}},\
  \bibinfo {editor} {edited by\ \bibinfo {editor} {\bibfnamefont
  {A.}~\bibnamefont {Mukherjee}}, \bibinfo {editor} {\bibfnamefont
  {M.}~\bibnamefont {Choudhury}}, \bibinfo {editor} {\bibfnamefont
  {F.}~\bibnamefont {Peruani}}, \bibinfo {editor} {\bibfnamefont
  {N.}~\bibnamefont {Ganguly}}, \ and\ \bibinfo {editor} {\bibfnamefont
  {B.}~\bibnamefont {Mitra}}}\ (\bibinfo  {publisher} {Springer New York},\
  \bibinfo {year} {2013})\ pp.\ \bibinfo {pages} {221--242}\BibitemShut
  {NoStop}%
\bibitem [{\citenamefont {Billeh}\ \emph {et~al.}(2014)\citenamefont {Billeh},
  \citenamefont {Schaub}, \citenamefont {Anastassiou}, \citenamefont
  {Barahona},\ and\ \citenamefont {Koch}}]{Billeh2014}%
  \BibitemOpen
  \bibfield  {author} {\bibinfo {author} {\bibfnamefont {Y.~N.}\ \bibnamefont
  {Billeh}}, \bibinfo {author} {\bibfnamefont {M.~T.}\ \bibnamefont {Schaub}},
  \bibinfo {author} {\bibfnamefont {C.~A.}\ \bibnamefont {Anastassiou}},
  \bibinfo {author} {\bibfnamefont {M.}~\bibnamefont {Barahona}}, \ and\
  \bibinfo {author} {\bibfnamefont {C.}~\bibnamefont {Koch}},\ }\href {\doibase
  10.1016/j.jneumeth.2014.08.011} {\bibfield  {journal} {\bibinfo  {journal}
  {Journal of Neuroscience Methods}\ }\textbf {\bibinfo {volume} {236}},\
  \bibinfo {pages} {92 } (\bibinfo {year} {2014})}\BibitemShut {NoStop}%
\bibitem [{\citenamefont {Rajan}\ and\ \citenamefont
  {Abbott}(2006)}]{Rajan2006}%
  \BibitemOpen
  \bibfield  {author} {\bibinfo {author} {\bibfnamefont {K.}~\bibnamefont
  {Rajan}}\ and\ \bibinfo {author} {\bibfnamefont {L.~F.}\ \bibnamefont
  {Abbott}},\ }\href {\doibase 10.1103/PhysRevLett.97.188104} {\bibfield
  {journal} {\bibinfo  {journal} {Physical Review Letters}\ }\textbf {\bibinfo
  {volume} {97}},\ \bibinfo {pages} {188104} (\bibinfo {year}
  {2006})}\BibitemShut {NoStop}%
\bibitem [{\citenamefont {Sommers}\ \emph {et~al.}(1988)\citenamefont
  {Sommers}, \citenamefont {Crisanti}, \citenamefont {Sompolinsky},\ and\
  \citenamefont {Stein}}]{Sommers1988}%
  \BibitemOpen
  \bibfield  {author} {\bibinfo {author} {\bibfnamefont {H.~J.}\ \bibnamefont
  {Sommers}}, \bibinfo {author} {\bibfnamefont {A.}~\bibnamefont {Crisanti}},
  \bibinfo {author} {\bibfnamefont {H.}~\bibnamefont {Sompolinsky}}, \ and\
  \bibinfo {author} {\bibfnamefont {Y.}~\bibnamefont {Stein}},\ }\href
  {\doibase 10.1103/PhysRevLett.60.1895} {\bibfield  {journal} {\bibinfo
  {journal} {Physical Review Letters}\ }\textbf {\bibinfo {volume} {60}},\
  \bibinfo {pages} {1895} (\bibinfo {year} {1988})}\BibitemShut {NoStop}%
\bibitem [{\citenamefont {Goldman}(2009)}]{Goldman2009}%
  \BibitemOpen
  \bibfield  {author} {\bibinfo {author} {\bibfnamefont {M.~S.}\ \bibnamefont
  {Goldman}},\ }\href {\doibase 10.1016/j.neuron.2008.12.012} {\bibfield
  {journal} {\bibinfo  {journal} {Neuron}\ }\textbf {\bibinfo {volume} {61}},\
  \bibinfo {pages} {621} (\bibinfo {year} {2009})}\BibitemShut {NoStop}%
\bibitem [{\citenamefont {Trefethen}\ and\ \citenamefont
  {Embree}(2005)}]{Trefethen2005}%
  \BibitemOpen
  \bibfield  {author} {\bibinfo {author} {\bibfnamefont {L.~N.}\ \bibnamefont
  {Trefethen}}\ and\ \bibinfo {author} {\bibfnamefont {M.}~\bibnamefont
  {Embree}},\ }\href@noop {} {\emph {\bibinfo {title} {{S}pectra and
  pseudospectra: the behavior of nonnormal matrices and operators}}}\ (\bibinfo
   {publisher} {Princeton University Press},\ \bibinfo {year}
  {2005})\BibitemShut {NoStop}%
\bibitem [{\citenamefont {Golub}\ and\ \citenamefont
  {Van~Loan}(1996)}]{Golub1996}%
  \BibitemOpen
  \bibfield  {author} {\bibinfo {author} {\bibfnamefont {H.}~\bibnamefont
  {Golub}}\ and\ \bibinfo {author} {\bibfnamefont {F.~C.}\ \bibnamefont
  {Van~Loan}},\ }\href@noop {} {\emph {\bibinfo {title} {{M}atrix
  {C}omputations}}},\ \bibinfo {edition} {3rd}\ ed.\ (\bibinfo  {publisher}
  {John Hopkins University Press},\ \bibinfo {address} {Baltimore and London},\
  \bibinfo {year} {1996})\BibitemShut {NoStop}%
\bibitem [{\citenamefont {Stewart}(2001)}]{Stewart2001}%
  \BibitemOpen
  \bibfield  {author} {\bibinfo {author} {\bibfnamefont {G.~W.}\ \bibnamefont
  {Stewart}},\ }\href@noop {} {\emph {\bibinfo {title} {{M}atrix {A}lgorithms
  {V}olume 2: {E}igensystems}}},\ Vol.~\bibinfo {volume} {2}\ (\bibinfo
  {publisher} {Siam},\ \bibinfo {year} {2001})\BibitemShut {NoStop}%
\bibitem [{\citenamefont {Rutishauser}\ \emph {et~al.}(2011)\citenamefont
  {Rutishauser}, \citenamefont {Douglas},\ and\ \citenamefont
  {Slotine}}]{Rutishauser2011}%
  \BibitemOpen
  \bibfield  {author} {\bibinfo {author} {\bibfnamefont {U.}~\bibnamefont
  {Rutishauser}}, \bibinfo {author} {\bibfnamefont {R.~J.}\ \bibnamefont
  {Douglas}}, \ and\ \bibinfo {author} {\bibfnamefont {J.-J.}\ \bibnamefont
  {Slotine}},\ }\href {\doibase {10.1162/NECO\_a\_00091}} {\bibfield  {journal}
  {\bibinfo  {journal} {Neural Computation}\ }\textbf {\bibinfo {volume}
  {23}},\ \bibinfo {pages} {735} (\bibinfo {year} {2011})}\BibitemShut
  {NoStop}%
\bibitem [{\citenamefont {Watts}\ and\ \citenamefont
  {Strogatz}(1998)}]{Watts1998}%
  \BibitemOpen
  \bibfield  {author} {\bibinfo {author} {\bibfnamefont {D.~J.}\ \bibnamefont
  {Watts}}\ and\ \bibinfo {author} {\bibfnamefont {S.~H.}\ \bibnamefont
  {Strogatz}},\ }\href {\doibase 10.1038/30918} {\bibfield  {journal} {\bibinfo
   {journal} {Nature}\ }\textbf {\bibinfo {volume} {393}},\ \bibinfo {pages}
  {440} (\bibinfo {year} {1998})}\BibitemShut {NoStop}%
\bibitem [{\citenamefont {Bullmore}\ and\ \citenamefont
  {Sporns}(2009)}]{Bullmore2009}%
  \BibitemOpen
  \bibfield  {author} {\bibinfo {author} {\bibfnamefont {E.}~\bibnamefont
  {Bullmore}}\ and\ \bibinfo {author} {\bibfnamefont {O.}~\bibnamefont
  {Sporns}},\ }\href {\doibase 10.1038/nrn2575} {\bibfield  {journal} {\bibinfo
   {journal} {Nature Reviews Neuroscience}\ }\textbf {\bibinfo {volume} {10}},\
  \bibinfo {pages} {186} (\bibinfo {year} {2009})}\BibitemShut {NoStop}%
\bibitem [{\citenamefont {Meunier}\ \emph {et~al.}(2010)\citenamefont
  {Meunier}, \citenamefont {Lambiotte},\ and\ \citenamefont
  {Bullmore}}]{Meunier2010}%
  \BibitemOpen
  \bibfield  {author} {\bibinfo {author} {\bibfnamefont {D.}~\bibnamefont
  {Meunier}}, \bibinfo {author} {\bibfnamefont {R.}~\bibnamefont {Lambiotte}},
  \ and\ \bibinfo {author} {\bibfnamefont {E.~T.}\ \bibnamefont {Bullmore}},\
  }\href {\doibase 10.3389/fnins.2010.00200} {\bibfield  {journal} {\bibinfo
  {journal} {Frontiers in Neuroscience}\ }\textbf {\bibinfo {volume} {4}}
  (\bibinfo {year} {2010}),\ 10.3389/fnins.2010.00200}\BibitemShut {NoStop}%
\bibitem [{\citenamefont {Barahona}\ and\ \citenamefont
  {Pecora}(2002)}]{Barahona2002}%
  \BibitemOpen
  \bibfield  {author} {\bibinfo {author} {\bibfnamefont {M.}~\bibnamefont
  {Barahona}}\ and\ \bibinfo {author} {\bibfnamefont {L.~M.}\ \bibnamefont
  {Pecora}},\ }\href {\doibase 10.1103/PhysRevLett.89.054101} {\bibfield
  {journal} {\bibinfo  {journal} {Physical Review Letters}\ }\textbf {\bibinfo
  {volume} {89}},\ \bibinfo {pages} {054101} (\bibinfo {year}
  {2002})}\BibitemShut {NoStop}%
\bibitem [{\citenamefont {Barabási}\ and\ \citenamefont
  {Albert}(1999)}]{Barabasi1999}%
  \BibitemOpen
  \bibfield  {author} {\bibinfo {author} {\bibfnamefont {A.-L.}\ \bibnamefont
  {Barabási}}\ and\ \bibinfo {author} {\bibfnamefont {R.}~\bibnamefont
  {Albert}},\ }\href {\doibase 10.1126/science.286.5439.509} {\bibfield
  {journal} {\bibinfo  {journal} {Science}\ }\textbf {\bibinfo {volume}
  {286}},\ \bibinfo {pages} {509} (\bibinfo {year} {1999})}\BibitemShut
  {NoStop}%
\bibitem [{\citenamefont {Ambrosingerson}\ \emph {et~al.}(1990)\citenamefont
  {Ambrosingerson}, \citenamefont {Granger},\ and\ \citenamefont
  {Lynch}}]{Ambrosingerson1990}%
  \BibitemOpen
  \bibfield  {author} {\bibinfo {author} {\bibfnamefont {J.}~\bibnamefont
  {Ambrosingerson}}, \bibinfo {author} {\bibfnamefont {R.}~\bibnamefont
  {Granger}}, \ and\ \bibinfo {author} {\bibfnamefont {G.}~\bibnamefont
  {Lynch}},\ }\href {\doibase 10.1126/science.2315702} {\bibfield  {journal}
  {\bibinfo  {journal} {Science}\ }\textbf {\bibinfo {volume} {247}},\ \bibinfo
  {pages} {1344} (\bibinfo {year} {1990})}\BibitemShut {NoStop}%
\bibitem [{\citenamefont {Laughlin}\ and\ \citenamefont
  {Sejnowski}(2003)}]{Laughlin2003}%
  \BibitemOpen
  \bibfield  {author} {\bibinfo {author} {\bibfnamefont {S.}~\bibnamefont
  {Laughlin}}\ and\ \bibinfo {author} {\bibfnamefont {T.}~\bibnamefont
  {Sejnowski}},\ }\href {\doibase {10.1126/science.1089662}} {\bibfield
  {journal} {\bibinfo  {journal} {Science}\ }\textbf {\bibinfo {volume}
  {301}},\ \bibinfo {pages} {1870} (\bibinfo {year} {2003})}\BibitemShut
  {NoStop}%
\bibitem [{\citenamefont {Kepecs}\ and\ \citenamefont
  {Fishell}(2014)}]{Kepecs2014}%
  \BibitemOpen
  \bibfield  {author} {\bibinfo {author} {\bibfnamefont {A.}~\bibnamefont
  {Kepecs}}\ and\ \bibinfo {author} {\bibfnamefont {G.}~\bibnamefont
  {Fishell}},\ }\href@noop {} {\bibfield  {journal} {\bibinfo  {journal}
  {Nature}\ }\textbf {\bibinfo {volume} {505}},\ \bibinfo {pages} {318–326}
  (\bibinfo {year} {2014})}\BibitemShut {NoStop}%
\bibitem [{\citenamefont {Roux}\ and\ \citenamefont
  {Buzs\'{a}ki}(2014)}]{Roux2014}%
  \BibitemOpen
  \bibfield  {author} {\bibinfo {author} {\bibfnamefont {L.}~\bibnamefont
  {Roux}}\ and\ \bibinfo {author} {\bibfnamefont {G.}~\bibnamefont
  {Buzs\'{a}ki}},\ }\href {\doibase
  http://dx.doi.org/10.1016/j.neuropharm.2014.09.011} {\bibfield  {journal}
  {\bibinfo  {journal} {Neuropharmacology}\ ,\ } (\bibinfo {year}
  {2014})}\BibitemShut {NoStop}%
\bibitem [{\citenamefont {Silberberg}\ and\ \citenamefont
  {Markram}(2007)}]{Silberberg2007}%
  \BibitemOpen
  \bibfield  {author} {\bibinfo {author} {\bibfnamefont {G.}~\bibnamefont
  {Silberberg}}\ and\ \bibinfo {author} {\bibfnamefont {H.}~\bibnamefont
  {Markram}},\ }\href {\doibase {10.1016/j.neuron.2007.02.012}} {\bibfield
  {journal} {\bibinfo  {journal} {Neuron}\ }\textbf {\bibinfo {volume} {53}},\
  \bibinfo {pages} {735} (\bibinfo {year} {2007})}\BibitemShut {NoStop}%
\bibitem [{\citenamefont {Xu}\ \emph {et~al.}(2013)\citenamefont {Xu},
  \citenamefont {Jeong}, \citenamefont {Tremblay},\ and\ \citenamefont
  {Rudy}}]{Xu2013}%
  \BibitemOpen
  \bibfield  {author} {\bibinfo {author} {\bibfnamefont {H.}~\bibnamefont
  {Xu}}, \bibinfo {author} {\bibfnamefont {H.-Y.}\ \bibnamefont {Jeong}},
  \bibinfo {author} {\bibfnamefont {R.}~\bibnamefont {Tremblay}}, \ and\
  \bibinfo {author} {\bibfnamefont {B.}~\bibnamefont {Rudy}},\ }\href {\doibase
  10.1016/j.neuron.2012.11.004} {\bibfield  {journal} {\bibinfo  {journal}
  {Neuron}\ }\textbf {\bibinfo {volume} {77}},\ \bibinfo {pages} {155}
  (\bibinfo {year} {2013})}\BibitemShut {NoStop}%
\bibitem [{\citenamefont {Lee}\ \emph {et~al.}(2013)\citenamefont {Lee},
  \citenamefont {Kruglikov}, \citenamefont {Huang}, \citenamefont {Fishell},\
  and\ \citenamefont {Rudy}}]{Lee2013}%
  \BibitemOpen
  \bibfield  {author} {\bibinfo {author} {\bibfnamefont {S.}~\bibnamefont
  {Lee}}, \bibinfo {author} {\bibfnamefont {I.}~\bibnamefont {Kruglikov}},
  \bibinfo {author} {\bibfnamefont {Z.~J.}\ \bibnamefont {Huang}}, \bibinfo
  {author} {\bibfnamefont {G.}~\bibnamefont {Fishell}}, \ and\ \bibinfo
  {author} {\bibfnamefont {B.}~\bibnamefont {Rudy}},\ }\href {\doibase
  10.1038/nn.3544} {\bibfield  {journal} {\bibinfo  {journal} {Nature
  Neuroscience}\ }\textbf {\bibinfo {volume} {16}},\ \bibinfo {pages} {1662}
  (\bibinfo {year} {2013})}\BibitemShut {NoStop}%
\bibitem [{\citenamefont {Pi}\ \emph {et~al.}(2013)\citenamefont {Pi},
  \citenamefont {Hangya}, \citenamefont {Kvitsiani}, \citenamefont {Sanders},
  \citenamefont {Huang},\ and\ \citenamefont {Kepecs}}]{Pi2013}%
  \BibitemOpen
  \bibfield  {author} {\bibinfo {author} {\bibfnamefont {H.-J.}\ \bibnamefont
  {Pi}}, \bibinfo {author} {\bibfnamefont {B.}~\bibnamefont {Hangya}}, \bibinfo
  {author} {\bibfnamefont {D.}~\bibnamefont {Kvitsiani}}, \bibinfo {author}
  {\bibfnamefont {J.~I.}\ \bibnamefont {Sanders}}, \bibinfo {author}
  {\bibfnamefont {Z.~J.}\ \bibnamefont {Huang}}, \ and\ \bibinfo {author}
  {\bibfnamefont {A.}~\bibnamefont {Kepecs}},\ }\href {\doibase
  10.1038/nature12676} {\bibfield  {journal} {\bibinfo  {journal} {Nature}\
  }\textbf {\bibinfo {volume} {503}},\ \bibinfo {pages} {521} (\bibinfo {year}
  {2013})}\BibitemShut {NoStop}%
\bibitem [{\citenamefont {Pfeffer}\ \emph {et~al.}(2013)\citenamefont
  {Pfeffer}, \citenamefont {Xue}, \citenamefont {He}, \citenamefont {Huang},\
  and\ \citenamefont {Scanziani}}]{Pfeffer2013}%
  \BibitemOpen
  \bibfield  {author} {\bibinfo {author} {\bibfnamefont {C.~K.}\ \bibnamefont
  {Pfeffer}}, \bibinfo {author} {\bibfnamefont {M.}~\bibnamefont {Xue}},
  \bibinfo {author} {\bibfnamefont {M.}~\bibnamefont {He}}, \bibinfo {author}
  {\bibfnamefont {Z.~J.}\ \bibnamefont {Huang}}, \ and\ \bibinfo {author}
  {\bibfnamefont {M.}~\bibnamefont {Scanziani}},\ }\href {\doibase
  10.1038/nn.3446} {\bibfield  {journal} {\bibinfo  {journal} {Nature
  Neuroscience}\ }\textbf {\bibinfo {volume} {16}},\ \bibinfo {pages} {1068}
  (\bibinfo {year} {2013})}\BibitemShut {NoStop}%
\bibitem [{\citenamefont {Niven}\ and\ \citenamefont
  {Laughlin}(2008)}]{Niven2008}%
  \BibitemOpen
  \bibfield  {author} {\bibinfo {author} {\bibfnamefont {J.~E.}\ \bibnamefont
  {Niven}}\ and\ \bibinfo {author} {\bibfnamefont {S.~B.}\ \bibnamefont
  {Laughlin}},\ }\href@noop {} {\bibfield  {journal} {\bibinfo  {journal}
  {Journal of Experimental Biology}\ }\textbf {\bibinfo {volume} {211}},\
  \bibinfo {pages} {1792} (\bibinfo {year} {2008})}\BibitemShut {NoStop}%
\bibitem [{\citenamefont {Larremore}\ \emph {et~al.}(2011)\citenamefont
  {Larremore}, \citenamefont {Shew},\ and\ \citenamefont
  {Restrepo}}]{Larremore2011}%
  \BibitemOpen
  \bibfield  {author} {\bibinfo {author} {\bibfnamefont {D.~B.}\ \bibnamefont
  {Larremore}}, \bibinfo {author} {\bibfnamefont {W.~L.}\ \bibnamefont {Shew}},
  \ and\ \bibinfo {author} {\bibfnamefont {J.~G.}\ \bibnamefont {Restrepo}},\
  }\href {\doibase 10.1103/PhysRevLett.106.058101} {\bibfield  {journal}
  {\bibinfo  {journal} {Physical Review Letters}\ }\textbf {\bibinfo {volume}
  {106}} (\bibinfo {year} {2011}),\ 10.1103/PhysRevLett.106.058101}\BibitemShut
  {NoStop}%
\bibitem [{\citenamefont {Beggs}\ and\ \citenamefont
  {Plenz}(2003)}]{Beggs2003}%
  \BibitemOpen
  \bibfield  {author} {\bibinfo {author} {\bibfnamefont {J.}~\bibnamefont
  {Beggs}}\ and\ \bibinfo {author} {\bibfnamefont {D.}~\bibnamefont {Plenz}},\
  }\href@noop {} {\bibfield  {journal} {\bibinfo  {journal} {Journal of
  Neuroscience}\ }\textbf {\bibinfo {volume} {23}},\ \bibinfo {pages} {11167}
  (\bibinfo {year} {2003})}\BibitemShut {NoStop}%
\bibitem [{\citenamefont {Kenet}\ \emph {et~al.}(2003)\citenamefont {Kenet},
  \citenamefont {Bibitchkov}, \citenamefont {Tsodyks}, \citenamefont
  {Grinvald},\ and\ \citenamefont {Arieli}}]{Kenet2003}%
  \BibitemOpen
  \bibfield  {author} {\bibinfo {author} {\bibfnamefont {T.}~\bibnamefont
  {Kenet}}, \bibinfo {author} {\bibfnamefont {D.}~\bibnamefont {Bibitchkov}},
  \bibinfo {author} {\bibfnamefont {M.}~\bibnamefont {Tsodyks}}, \bibinfo
  {author} {\bibfnamefont {A.}~\bibnamefont {Grinvald}}, \ and\ \bibinfo
  {author} {\bibfnamefont {A.}~\bibnamefont {Arieli}},\ }\href {\doibase
  10.1038/nature02078} {\bibfield  {journal} {\bibinfo  {journal} {Nature}\
  }\textbf {\bibinfo {volume} {425}},\ \bibinfo {pages} {954} (\bibinfo {year}
  {2003})}\BibitemShut {NoStop}%
\bibitem [{\citenamefont {Ben-Yishai}\ \emph {et~al.}(1995)\citenamefont
  {Ben-Yishai}, \citenamefont {Bar-Or},\ and\ \citenamefont
  {Sompolinsky}}]{BenYishai1995}%
  \BibitemOpen
  \bibfield  {author} {\bibinfo {author} {\bibfnamefont {R.}~\bibnamefont
  {Ben-Yishai}}, \bibinfo {author} {\bibfnamefont {R.}~\bibnamefont {Bar-Or}},
  \ and\ \bibinfo {author} {\bibfnamefont {H.}~\bibnamefont {Sompolinsky}},\
  }\href {\doibase 10.1073/pnas.92.9.3844} {\bibfield  {journal} {\bibinfo
  {journal} {Proceedings of the National Academy of Sciences of the United
  States of America}\ }\textbf {\bibinfo {volume} {92}},\ \bibinfo {pages}
  {3844} (\bibinfo {year} {1995})}\BibitemShut {NoStop}%
\bibitem [{\citenamefont {Somers}\ \emph {et~al.}(1995)\citenamefont {Somers},
  \citenamefont {Nelson},\ and\ \citenamefont {Sur}}]{Somers1995}%
  \BibitemOpen
  \bibfield  {author} {\bibinfo {author} {\bibfnamefont {D.}~\bibnamefont
  {Somers}}, \bibinfo {author} {\bibfnamefont {S.}~\bibnamefont {Nelson}}, \
  and\ \bibinfo {author} {\bibfnamefont {M.}~\bibnamefont {Sur}},\ }\href@noop
  {} {\bibfield  {journal} {\bibinfo  {journal} {Journal of Neuroscience}\
  }\textbf {\bibinfo {volume} {15}},\ \bibinfo {pages} {5448} (\bibinfo {year}
  {1995})}\BibitemShut {NoStop}%
\bibitem [{\citenamefont {Ernst}\ \emph {et~al.}(2001)\citenamefont {Ernst},
  \citenamefont {Pawelzik}, \citenamefont {Sahar-Pikielny},\ and\ \citenamefont
  {Tsodyks}}]{Ernst2001}%
  \BibitemOpen
  \bibfield  {author} {\bibinfo {author} {\bibfnamefont {U.}~\bibnamefont
  {Ernst}}, \bibinfo {author} {\bibfnamefont {K.}~\bibnamefont {Pawelzik}},
  \bibinfo {author} {\bibfnamefont {C.}~\bibnamefont {Sahar-Pikielny}}, \ and\
  \bibinfo {author} {\bibfnamefont {M.}~\bibnamefont {Tsodyks}},\ }\href
  {\doibase 10.1038/86089} {\bibfield  {journal} {\bibinfo  {journal} {Nature
  Neuroscience}\ }\textbf {\bibinfo {volume} {4}},\ \bibinfo {pages} {431}
  (\bibinfo {year} {2001})}\BibitemShut {NoStop}%
\bibitem [{\citenamefont {Sasaki}\ \emph {et~al.}(2007)\citenamefont {Sasaki},
  \citenamefont {Matsuki},\ and\ \citenamefont {Ikegaya}}]{Sasaki2007}%
  \BibitemOpen
  \bibfield  {author} {\bibinfo {author} {\bibfnamefont {T.}~\bibnamefont
  {Sasaki}}, \bibinfo {author} {\bibfnamefont {N.}~\bibnamefont {Matsuki}}, \
  and\ \bibinfo {author} {\bibfnamefont {Y.}~\bibnamefont {Ikegaya}},\ }\href
  {\doibase 10.1523/JNEUROSCI.4514-06.2007} {\bibfield  {journal} {\bibinfo
  {journal} {Journal of Neuroscience}\ }\textbf {\bibinfo {volume} {27}},\
  \bibinfo {pages} {517} (\bibinfo {year} {2007})}\BibitemShut {NoStop}%
\bibitem [{\citenamefont {Litwin-Kumar}\ and\ \citenamefont
  {Doiron}(2014)}]{Litwin-Kumar2014}%
  \BibitemOpen
  \bibfield  {author} {\bibinfo {author} {\bibfnamefont {A.}~\bibnamefont
  {Litwin-Kumar}}\ and\ \bibinfo {author} {\bibfnamefont {B.}~\bibnamefont
  {Doiron}},\ }\href@noop {} {\bibfield  {journal} {\bibinfo  {journal} {Nature
  Communications}\ }\textbf {\bibinfo {volume} {5}} (\bibinfo {year}
  {2014})}\BibitemShut {NoStop}%
\bibitem [{\citenamefont {Harris}\ \emph {et~al.}(2003)\citenamefont {Harris},
  \citenamefont {Csicsvari}, \citenamefont {Hirase}, \citenamefont {Dragoi},\
  and\ \citenamefont {Buzsaki}}]{Harris2003}%
  \BibitemOpen
  \bibfield  {author} {\bibinfo {author} {\bibfnamefont {K.}~\bibnamefont
  {Harris}}, \bibinfo {author} {\bibfnamefont {J.}~\bibnamefont {Csicsvari}},
  \bibinfo {author} {\bibfnamefont {H.}~\bibnamefont {Hirase}}, \bibinfo
  {author} {\bibfnamefont {G.}~\bibnamefont {Dragoi}}, \ and\ \bibinfo {author}
  {\bibfnamefont {G.}~\bibnamefont {Buzsaki}},\ }\href {\doibase
  10.1038/nature01834} {\bibfield  {journal} {\bibinfo  {journal} {Nature}\
  }\textbf {\bibinfo {volume} {424}},\ \bibinfo {pages} {552} (\bibinfo {year}
  {2003})}\BibitemShut {NoStop}%
\bibitem [{\citenamefont {Foster}\ and\ \citenamefont
  {Wilson}(2006)}]{Foster2006}%
  \BibitemOpen
  \bibfield  {author} {\bibinfo {author} {\bibfnamefont {D.}~\bibnamefont
  {Foster}}\ and\ \bibinfo {author} {\bibfnamefont {M.}~\bibnamefont
  {Wilson}},\ }\href {\doibase 10.1038/nature04587} {\bibfield  {journal}
  {\bibinfo  {journal} {Nature}\ }\textbf {\bibinfo {volume} {440}},\ \bibinfo
  {pages} {680} (\bibinfo {year} {2006})}\BibitemShut {NoStop}%
\bibitem [{\citenamefont {Riehle}\ \emph {et~al.}(1997)\citenamefont {Riehle},
  \citenamefont {Grun}, \citenamefont {Diesmann},\ and\ \citenamefont
  {Aertsen}}]{Riehle1997}%
  \BibitemOpen
  \bibfield  {author} {\bibinfo {author} {\bibfnamefont {A.}~\bibnamefont
  {Riehle}}, \bibinfo {author} {\bibfnamefont {S.}~\bibnamefont {Grun}},
  \bibinfo {author} {\bibfnamefont {M.}~\bibnamefont {Diesmann}}, \ and\
  \bibinfo {author} {\bibfnamefont {A.}~\bibnamefont {Aertsen}},\ }\href
  {\doibase 10.1126/science.278.5345.1950} {\bibfield  {journal} {\bibinfo
  {journal} {Science}\ }\textbf {\bibinfo {volume} {278}},\ \bibinfo {pages}
  {1950} (\bibinfo {year} {1997})}\BibitemShut {NoStop}%
\bibitem [{\citenamefont {Hahnloser}\ \emph {et~al.}(2002)\citenamefont
  {Hahnloser}, \citenamefont {Kozhevnikov},\ and\ \citenamefont
  {Fee}}]{Hahnloser2002}%
  \BibitemOpen
  \bibfield  {author} {\bibinfo {author} {\bibfnamefont {R.}~\bibnamefont
  {Hahnloser}}, \bibinfo {author} {\bibfnamefont {A.}~\bibnamefont
  {Kozhevnikov}}, \ and\ \bibinfo {author} {\bibfnamefont {M.}~\bibnamefont
  {Fee}},\ }\href {\doibase 10.1038/nature00974} {\bibfield  {journal}
  {\bibinfo  {journal} {Nature}\ }\textbf {\bibinfo {volume} {419}},\ \bibinfo
  {pages} {65} (\bibinfo {year} {2002})}\BibitemShut {NoStop}%
\bibitem [{\citenamefont {Rokem}\ \emph {et~al.}(2006)\citenamefont {Rokem},
  \citenamefont {Watzl}, \citenamefont {Gollisch}, \citenamefont {Stemmler},
  \citenamefont {Herz},\ and\ \citenamefont {Samengo}}]{Rokem2006}%
  \BibitemOpen
  \bibfield  {author} {\bibinfo {author} {\bibfnamefont {A.}~\bibnamefont
  {Rokem}}, \bibinfo {author} {\bibfnamefont {S.}~\bibnamefont {Watzl}},
  \bibinfo {author} {\bibfnamefont {T.}~\bibnamefont {Gollisch}}, \bibinfo
  {author} {\bibfnamefont {M.}~\bibnamefont {Stemmler}}, \bibinfo {author}
  {\bibfnamefont {A.}~\bibnamefont {Herz}}, \ and\ \bibinfo {author}
  {\bibfnamefont {I.}~\bibnamefont {Samengo}},\ }\href {\doibase
  10.1152/jn.00891.2005} {\bibfield  {journal} {\bibinfo  {journal} {Journal of
  Neurophysiology}\ }\textbf {\bibinfo {volume} {95}},\ \bibinfo {pages} {2541}
  (\bibinfo {year} {2006})}\BibitemShut {NoStop}%
\bibitem [{\citenamefont {Hromadka}\ \emph {et~al.}(2008)\citenamefont
  {Hromadka}, \citenamefont {DeWeese},\ and\ \citenamefont
  {Zador}}]{Hromadka2008}%
  \BibitemOpen
  \bibfield  {author} {\bibinfo {author} {\bibfnamefont {T.}~\bibnamefont
  {Hromadka}}, \bibinfo {author} {\bibfnamefont {M.~R.}\ \bibnamefont
  {DeWeese}}, \ and\ \bibinfo {author} {\bibfnamefont {A.~M.}\ \bibnamefont
  {Zador}},\ }\href {\doibase 10.1371/journal.pbio.0060016} {\bibfield
  {journal} {\bibinfo  {journal} {PLoS Biology}\ }\textbf {\bibinfo {volume}
  {6}},\ \bibinfo {pages} {124} (\bibinfo {year} {2008})}\BibitemShut {NoStop}%
\bibitem [{\citenamefont {Buzsaki}\ and\ \citenamefont
  {Mizuseki}(2014)}]{Buzsaki2014}%
  \BibitemOpen
  \bibfield  {author} {\bibinfo {author} {\bibfnamefont {G.}~\bibnamefont
  {Buzsaki}}\ and\ \bibinfo {author} {\bibfnamefont {K.}~\bibnamefont
  {Mizuseki}},\ }\href {\doibase 10.1038/nrn3687} {\bibfield  {journal}
  {\bibinfo  {journal} {Nature Reviews Neuroscience}\ }\textbf {\bibinfo
  {volume} {15}},\ \bibinfo {pages} {264} (\bibinfo {year} {2014})}\BibitemShut
  {NoStop}%
\bibitem [{\citenamefont {Oh}\ \emph {et~al.}(2014)\citenamefont {Oh},
  \citenamefont {Harris}, \citenamefont {Ng}, \citenamefont {Winslow},
  \citenamefont {Cain}, \citenamefont {Mihalas}, \citenamefont {Wang},
  \citenamefont {Lau}, \citenamefont {Kuan}, \citenamefont {Henry},
  \citenamefont {Mortrud}, \citenamefont {Ouellette}, \citenamefont {Nguyen},
  \citenamefont {Sorensen}, \citenamefont {Slaughterbeck}, \citenamefont
  {Wakeman}, \citenamefont {Li}, \citenamefont {Feng}, \citenamefont {Ho},
  \citenamefont {Nicholas}, \citenamefont {Hirokawa}, \citenamefont {Bohn},
  \citenamefont {Joines}, \citenamefont {Peng}, \citenamefont {Hawrylycz},
  \citenamefont {Phillips}, \citenamefont {Hohmann}, \citenamefont {Wohnoutka},
  \citenamefont {Koch}, \citenamefont {Bernard}, \citenamefont {Dang},
  \citenamefont {Jones}, \citenamefont {Zeng},\ and\ \citenamefont
  {Gerfen}}]{Oh2014}%
  \BibitemOpen
  \bibfield  {author} {\bibinfo {author} {\bibfnamefont {S.~W.}\ \bibnamefont
  {Oh}}, \bibinfo {author} {\bibfnamefont {J.~A.}\ \bibnamefont {Harris}},
  \bibinfo {author} {\bibfnamefont {L.}~\bibnamefont {Ng}}, \bibinfo {author}
  {\bibfnamefont {B.}~\bibnamefont {Winslow}}, \bibinfo {author} {\bibfnamefont
  {N.}~\bibnamefont {Cain}}, \bibinfo {author} {\bibfnamefont {S.}~\bibnamefont
  {Mihalas}}, \bibinfo {author} {\bibfnamefont {Q.}~\bibnamefont {Wang}},
  \bibinfo {author} {\bibfnamefont {C.}~\bibnamefont {Lau}}, \bibinfo {author}
  {\bibfnamefont {L.}~\bibnamefont {Kuan}}, \bibinfo {author} {\bibfnamefont
  {A.~M.}\ \bibnamefont {Henry}}, \bibinfo {author} {\bibfnamefont {M.~T.}\
  \bibnamefont {Mortrud}}, \bibinfo {author} {\bibfnamefont {B.}~\bibnamefont
  {Ouellette}}, \bibinfo {author} {\bibfnamefont {T.~N.}\ \bibnamefont
  {Nguyen}}, \bibinfo {author} {\bibfnamefont {S.~A.}\ \bibnamefont
  {Sorensen}}, \bibinfo {author} {\bibfnamefont {C.~R.}\ \bibnamefont
  {Slaughterbeck}}, \bibinfo {author} {\bibfnamefont {W.}~\bibnamefont
  {Wakeman}}, \bibinfo {author} {\bibfnamefont {Y.}~\bibnamefont {Li}},
  \bibinfo {author} {\bibfnamefont {D.}~\bibnamefont {Feng}}, \bibinfo {author}
  {\bibfnamefont {A.}~\bibnamefont {Ho}}, \bibinfo {author} {\bibfnamefont
  {E.}~\bibnamefont {Nicholas}}, \bibinfo {author} {\bibfnamefont {K.~E.}\
  \bibnamefont {Hirokawa}}, \bibinfo {author} {\bibfnamefont {P.}~\bibnamefont
  {Bohn}}, \bibinfo {author} {\bibfnamefont {K.~M.}\ \bibnamefont {Joines}},
  \bibinfo {author} {\bibfnamefont {H.}~\bibnamefont {Peng}}, \bibinfo {author}
  {\bibfnamefont {M.~J.}\ \bibnamefont {Hawrylycz}}, \bibinfo {author}
  {\bibfnamefont {J.~W.}\ \bibnamefont {Phillips}}, \bibinfo {author}
  {\bibfnamefont {J.~G.}\ \bibnamefont {Hohmann}}, \bibinfo {author}
  {\bibfnamefont {P.}~\bibnamefont {Wohnoutka}}, \bibinfo {author}
  {\bibfnamefont {C.}~\bibnamefont {Koch}}, \bibinfo {author} {\bibfnamefont
  {A.}~\bibnamefont {Bernard}}, \bibinfo {author} {\bibfnamefont
  {C.}~\bibnamefont {Dang}}, \bibinfo {author} {\bibfnamefont {A.~R.}\
  \bibnamefont {Jones}}, \bibinfo {author} {\bibfnamefont {H.}~\bibnamefont
  {Zeng}}, \ and\ \bibinfo {author} {\bibfnamefont {C.~R.}\ \bibnamefont
  {Gerfen}},\ }\href {\doibase {10.1038/nature13186}} {\bibfield  {journal}
  {\bibinfo  {journal} {Nature}\ }\textbf {\bibinfo {volume} {508}},\ \bibinfo
  {pages} {207} (\bibinfo {year} {2014})}\BibitemShut {NoStop}%
\bibitem [{\citenamefont {Clopath}\ \emph {et~al.}(2010)\citenamefont
  {Clopath}, \citenamefont {Buesing}, \citenamefont {Vasilaki},\ and\
  \citenamefont {Gerstner}}]{Clopath2010}%
  \BibitemOpen
  \bibfield  {author} {\bibinfo {author} {\bibfnamefont {C.}~\bibnamefont
  {Clopath}}, \bibinfo {author} {\bibfnamefont {L.}~\bibnamefont {Buesing}},
  \bibinfo {author} {\bibfnamefont {E.}~\bibnamefont {Vasilaki}}, \ and\
  \bibinfo {author} {\bibfnamefont {W.}~\bibnamefont {Gerstner}},\ }\href
  {\doibase 10.1038/nn.2479} {\bibfield  {journal} {\bibinfo  {journal} {Nature
  Neuroscience}\ }\textbf {\bibinfo {volume} {13}},\ \bibinfo {pages} {344}
  (\bibinfo {year} {2010})}\BibitemShut {NoStop}%
\bibitem [{\citenamefont {Reimann}\ \emph {et~al.}(2013)\citenamefont
  {Reimann}, \citenamefont {Anastassiou}, \citenamefont {Perin}, \citenamefont
  {Hill}, \citenamefont {Markram},\ and\ \citenamefont {Koch}}]{Reimann2013}%
  \BibitemOpen
  \bibfield  {author} {\bibinfo {author} {\bibfnamefont {M.~W.}\ \bibnamefont
  {Reimann}}, \bibinfo {author} {\bibfnamefont {C.~A.}\ \bibnamefont
  {Anastassiou}}, \bibinfo {author} {\bibfnamefont {R.}~\bibnamefont {Perin}},
  \bibinfo {author} {\bibfnamefont {S.~L.}\ \bibnamefont {Hill}}, \bibinfo
  {author} {\bibfnamefont {H.}~\bibnamefont {Markram}}, \ and\ \bibinfo
  {author} {\bibfnamefont {C.}~\bibnamefont {Koch}},\ }\href {\doibase
  10.1016/j.neuron.2013.05.023} {\bibfield  {journal} {\bibinfo  {journal}
  {Neuron}\ }\textbf {\bibinfo {volume} {79}},\ \bibinfo {pages} {375}
  (\bibinfo {year} {2013})}\BibitemShut {NoStop}%
\bibitem [{\citenamefont {Batagelj}\ and\ \citenamefont
  {Brandes}(2005)}]{Batagelj2005}%
  \BibitemOpen
  \bibfield  {author} {\bibinfo {author} {\bibfnamefont {V.}~\bibnamefont
  {Batagelj}}\ and\ \bibinfo {author} {\bibfnamefont {U.}~\bibnamefont
  {Brandes}},\ }\href {\doibase 10.1103/PhysRevE.71.036113} {\bibfield
  {journal} {\bibinfo  {journal} {Phys. Rev. E}\ }\textbf {\bibinfo {volume}
  {71}},\ \bibinfo {pages} {036113} (\bibinfo {year} {2005})}\BibitemShut
  {NoStop}%
\end{thebibliography}%

\renewcommand{\thefigure}{S\arabic{figure}}
\setcounter{figure}{0}
\appendix
\pagebreak
\cleardoublepage

\section{Supplementary Information -- Emergence of slow-switching assemblies in structured neuronal networks}

\begin{figure}[htb!]
 \centering
 \includegraphics{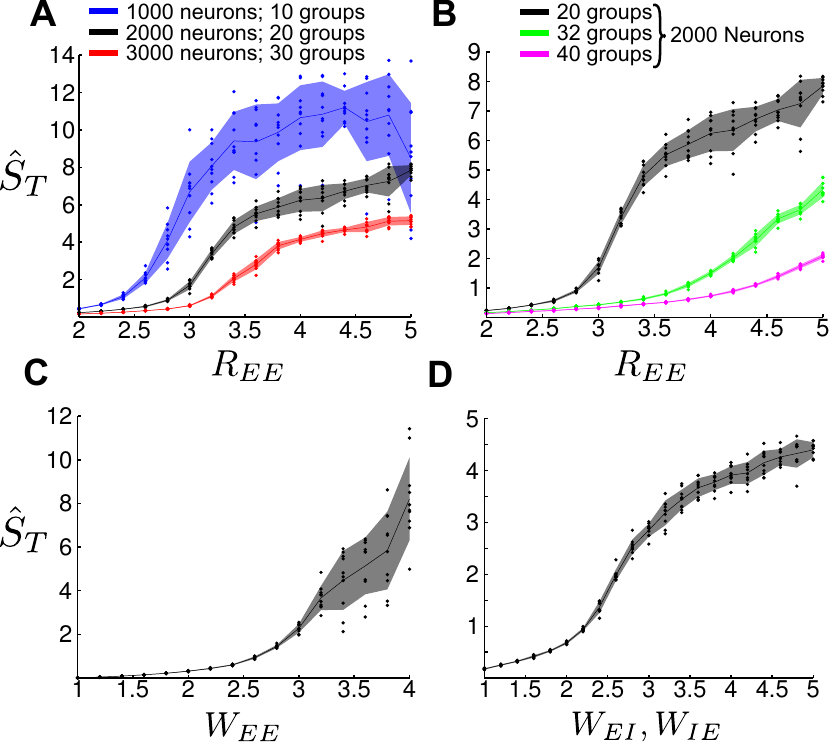}
 \caption{Spike rate variability across time, as a function of topological changes of the network. 
\textbf{A}~Clustering Strength $R_{EE}$ vs spike rate variability across time $\hat S_T$ for varying network sizes with a fixed assembly size.  Compare to Fig.~2A of the main text.
\textbf{B}~Clustering Strength $R_{EE}$ vs spike rate variability across time $\hat S_T$ for a fixed network size with varying assembly sizes. Compare to Fig.~2B of the main text.
\textbf{C}~Weight Clustering $W_{EE}$ vs spike rate variability across time $\hat S_T$ for a LIF network with 2000 neurons (20 groups; topological clustering $R_{EE}=1$). Compare to Fig.~4 of the main text.
\textbf{D}~Spike rate variability across time $\hat S_T$ as a function of ${W_{EI} = W_{IE}}$ for a network with excitatory to inhibitory feedback loops. Compare to Fig.~8 of the main text. In all plots dots denote raw data from simulations; line: mean, shading: standard deviation.}
 \label{fig:S1}
\end{figure}

\begin{figure}[htb!]
 \centering
 \includegraphics{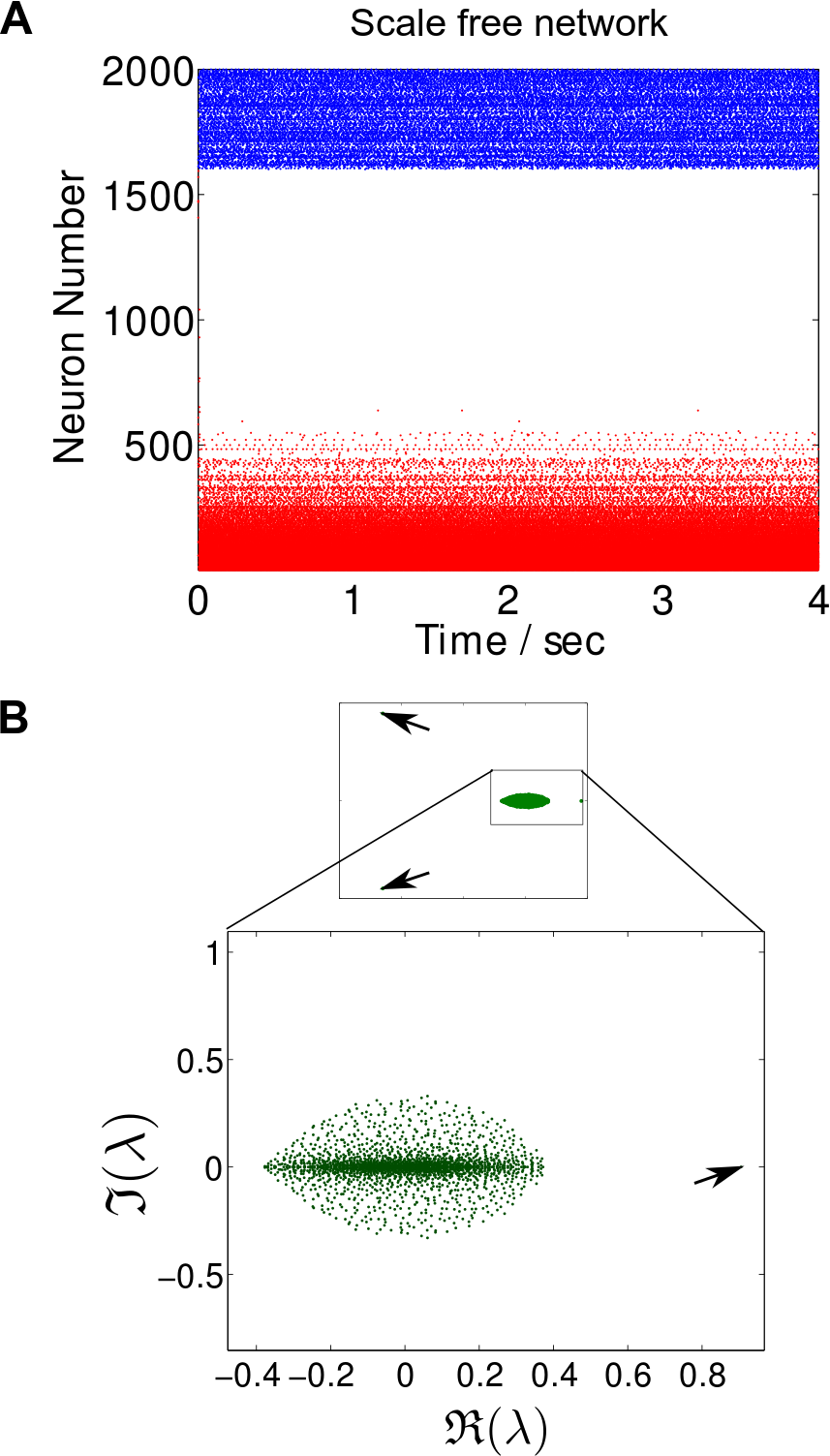}
 \caption{Network dynamics for a scale-free network. 
  \textbf{A}~Raster plot for an example scale-free network. Excitatory neurons are ordered according to their (expected) degree: high (bottom) to low (top). Note that the spiking activity is largely concentrated around the hub (which due to assortativity is connected to other nodes of high degree), and there are no distinguishable groups or  switching behavior.
  \textbf{B} Spectrum of the simulated scale-free network. There is only one eigenvalue separated from the main bulk of the spectrum due to the presence of the large hub in the network.}
 \label{fig:S3}
\end{figure}

\end{document}